\documentclass{emulateapj}
\shorttitle{Bulk Properties of Nearby Clusters}
\shortauthors{O'Hara et al.}

\newcommand{\Chandra}{{\it Chandra}}
\newcommand{\ROSAT}{{\it ROSAT}}

\def\Micm{$M_{\rm ICM}$}

\def\Micmfive{$M_{\rm ICM}(<$$r_{500})$}
\def\Micmtfive{$M_{\rm ICM}(<$$r_{2500})$}
\def\L2500{$L_{2500}$}
\def\Lfive{$L_{500}$}
\def\LNIR{$L_{\rm NIR}$}
\def\LNIRfive{$L_{\rm NIR}(<$$r_{500})$}
\def\M2500{$M_{2500}$}
\def\rtfive{$r_{2500}$}
\def\rd{$r_\Delta$}
\def\rfive{$r_{500}$}
\def\rtwo{$r_{200}$}
\def\AI{$A_{I}$}
\def\RI{$R_{I}$}
\def\LX{$L_{\rm X}$}
\def\LXCS{$L_{\rm XCS}$}
\def\LXCSfive{$L_{\rm XCS}(<$$r_{500})$}
\def\LXfive{$L_{\rm X}(<$$r_{500})$}
\def\LXtfive{$L_{\rm X}(<$$r_{2500})$}
\def\sint{$\sigma_{\rm int}$}
\def\sraw{$\sigma_{\rm raw}$}
\def\Rthree{$R_{3\times10^{-14}}$}
\def\Rone{$R_{1.5\times10^{-13}}$}
\def\Tx{$T_{\rm X}$}

\def\ps{s$^{-1}$}
\def\Pone{$P_1^{(pk)}/P_0^{(pk)}$}
\def\Ptwo{$P_2/P_0$}

\def\Io{$I_0$}
\def\OTI{$\mathcal{O}$--$T_{\rm X}$--$I_0$}
\def\OT{$\mathcal{O}$--$T_{\rm X}$}

\def\myputfigure#1#2#3#4#5%
{\vskip#5pt\makebox[0pt]{\hskip#2in
\includegraphics[width=#3\textwidth]{#1}}\vskip#4pt\hfill}
\newenvironment{inlinefigure}{
\def\@captype{figure}
\noindent\begin{minipage}{0.999\linewidth}\begin{center}}
{\end{center}\end{minipage}\smallskip}

\begin{document}

\submitted{Submitted to ApJ August 2, 2005}


\title{Effects of Mergers and Core Structure 
on the Bulk Properties of \\ Nearby Galaxy Clusters}

\author{Timothy B. O'Hara,\altaffilmark{1,2} Joseph J. Mohr,\altaffilmark{1,2}, John J. Bialek,\altaffilmark{3} \& August E. Evrard\altaffilmark{3,4}}

\altaffiltext{1}{Department of Physics, University of Illinois,
1110 West Green St, Urbana, IL 61801; tbohara@astro.uiuc.edu, jmohr@uiuc.edu}
\altaffiltext{2}{Department of Astronomy, University of Illinois,
1002 West Green St, Urbana, IL 61801}
\altaffiltext{3}{Department of Physics, 1049 Randall Laboratory, University of Michigan at Ann Arbor, Ann Arbor, MI 48109}
\altaffiltext{4}{Department of Astronomy, Dennison Building, University of Michigan at Ann Arbor, Ann Arbor, MI 48109}

\begin{abstract}

We use X-ray morphological measurements and the scatter of clusters about observed and simulated scaling relations to examine the impact of merging and core-related phenomena on the structure of galaxy clusters.  We use a range of X-ray and near-infrared scaling relations; all observed scaling relations constructed from emission-weighted mean temperature and intracluster medium mass, X-ray luminosity, isophotal size, and near-IR luminosity show a separation between clusters identified as cool core (CC) and those identified as non-cool core (NCC). We attribute this partially to a simple temperature bias in CC clusters, and partially to other cool core-related structural changes. Scaling relations constructed from observables that are largely independent of core structure show smaller separation between CC and NCC populations.
We attempt to minimize cool core-related separation in scaling relations via two methods: by applying a uniform scale factor to CC cluster temperatures and determining the scale factor for each relation that minimizes the separation between CC and NCC populations, and by introducing cluster central surface brightness as a third parameter in observable--temperature scaling relations.  We find an average temperature bias factor of $1.07 \pm 0.02$ between the CC and NCC populations;  the three parameter approach reduces scatter in scaling relations more than a simple CC temperature scaling.

We examine the scatter about the best-fit observable--temperature--brightness scaling relations, and compare the intrinsic scatter within subsamples split by CC/NCC and four different morphological merger indicators.  CC clusters and clusters with {\it less} substructure generally exhibit higher scatter about scaling relations.  The larger structural variations in CC clusters are present well outside the core, suggesting that a process more global than core radiative instability is at work.  Simulations without cooling mechanisms also show no  correlation between substructure and larger scatter about scaling relations, indicating that any merger-related scatter increases are subtle. Taken together, the observational and simulation results indicate that cool core related phenomena---not merging processes---are the primary contributor to scatter in scaling relations.  Our analysis does not appear to support the scenario in which clusters evolve cool cores over time unless they experience major mergers.

\end{abstract}


\section{Introduction}
\label{sec:intro}

Galaxy clusters provide a setting for exploring the composition of the universe on large scales, and for studying the growth of structure.  X-ray and Sunyaev-Zel'dovich effect surveys of large numbers of clusters will soon be used to study the nature of the dark energy. It is thus vitally important to understand cluster structure and its connections to observable, bulk properties of clusters.

Clusters exhibit strikingly regular power law scaling relations between such properties as emission-weighted mean temperature, intracluster medium (ICM) mass, binding mass, X-ray luminosity, isophotal size, and near-IR light \citep[e.g., ][]{mohr97a,mohr99,vikhlinin02,lin03b}. For relaxed, isolated systems this is expected, although even for isolated systems there should be some scatter at a given average temperature due to variations in formation epoch, gas fraction, and galaxy and star formation history.

However, clusters are in fact young, dynamic systems. There is evidence, much of it based on quantitative substructure measures, for merging in a significant fraction ($>$50\%) of nearby clusters \citep[e.g., ][]{geller82,dressler88,mohr95,buote96}. The advent of high-resolution X-ray instruments such as \Chandra\  has made observation of the complex hydrodynamical structure of merging clusters possible, revealing and permitting detailed measurements of ICM properties around features such as ``cold fronts" \citep{vikhlinin01}.

This dual nature of clusters---frequent merging on the one hand, and regular scaling relations on the other---is puzzling. The extreme energetics of merger events, in which $\sim10^{63}$ erg of kinetic energy can be thermalized and cluster structure greatly disrupted, would suggest that cluster properties should not be correlated in such a simple way.  One might expect to find a statistical correlation between the deviation of a particular cluster from a scaling relation and the substructure---an indication of merger activity---in that cluster.

Tight scaling relations are observed in simulations of clusters that evolve in a cosmological context \citep[e.g.,][]{evrard96, bryan98, bialek01}. However, simulations of mergers of isolated, relaxed clusters suggest that merging clusters should exhibit massive boosts in cluster bulk parameters such as temperature and X-ray luminosity \citep{ricker01,randall02}. While the simulations indicate that these boosts are correlated, it seems likely that a cluster involved in a major merger would stand out from scaling relations constructed from several observables simultaneously. 

If the cool cores found in a large fraction of the cluster population are an expected outcome of cluster relaxation in the absence of merger activity \citep[e.g.,][]{ota05}, then we may expect ``relaxed" clusters to display less structural variation than clusters recently involved in mergers. If, on the other hand, cool core structure is linked to more fundamental properties of individual clusters, such as early-time entropy injection \citep{mccarthy04}, then merger-related effects may be overshadowed. There is presently no consensus as to whether merger-related effects dominate the scatter in scaling relations \citep[e.g.,][]{smith05}, or are a minority contribution \citep[e.g.,][]{balogh05}.

In this paper we study multiple scaling relations. One may envision surfaces of clusters in a hyperplane constructed from several observables, and imagine studying deviations from these relations in multiple dimensions. We begin by examining only two parameters at a time, largely due to the greater ease of visualizing and understanding the deviations.  Later we combine these two-observable analyses to examine cluster departures from the population in the much higher-dimensionality space.  Our hope is that by examining nine different X-ray and near-infrared observables,  each of which represents a different integral over the structure of the ICM and galaxies, we will be able to discern even modest structural deviations and probe their relationship to merging and other cluster phenomena.

If, as expected, cluster mergers perturb crude observables---for example, producing erroneously high estimates of cluster mass---then this must be taken into account in surveys using clusters to study cosmology \citep[e.g.][]{haiman01,hu03b,majumdar03,majumdar04,lima05}. It is thus of great importance to understand the true effects of merging on cluster bulk properties. Positions on scaling relations have been examined for a few individual merging clusters, e.g., Cl J0152.7--1357 \citep{maughan03} and A2319 \citep{ohara04}, with no significant deviations found. Clearly, however, analysis of a larger sample of clusters is needed to make meaningful statements about the relationship between merger signatures and bulk properties. Such an undertaking is the subject of this paper. 

Scaling relations have already been used to study cluster structure; for example, by examining the the effects of structure on the slope of the relations \citep[e.g.,][]{evrard96,cavaliere97,mohr97a,bialek01}.
The trend in recent X-ray studies of clusters, however, has been toward detailed studies of clusters using high-resolution instruments such as \Chandra, and correspondingly detailed simulations \citep[e.g.,][]{ricker01,nagai03,onuora03}. These studies have uncovered many surprising aspects of cluster structure. There is still a need, though, for a clearer picture of the effects of merging on the entire cluster population. An analysis of populations of simulated clusters suggests that it will not be possible to isolate ``undisturbed"  or ``relaxed'' clusters in large samples over a range of redshifts, and so a better understanding of merger effects on cluster bulk properties and on the general population is required. 

\tabletypesize{\tiny}
\begin{deluxetable*}{p{1.2cm}cccccccccccc}

\tablewidth{0pt}
\tabletypesize{\tiny}
\tablecaption{PSPC Sample Information}
\tablehead{
Cluster & $z$ &
\colhead{\Tx} & \colhead{\Micmfive} & \colhead{\Micmtfive} & \colhead{$R_{3{\rm E}-14}$} & \colhead{$R_{1.5{\rm E}-13}$} & \colhead{\Lfive} & \colhead{\L2500} & \colhead{\LXCS} & \colhead{\LNIRfive} & \colhead{CC} & \colhead{\Tx} \\
 &  & \colhead{(keV)} &
\colhead{($h_{70}^{-5/2}$ $M_\odot$)} &
\colhead{($h_{70}^{-5/2}$ $M_\odot$)} &
\colhead{($h_{70}^{-1}$ Mpc)} &
\colhead{($h_{70}^{-1}$ Mpc)} &
\colhead{($h_{70}^{-2}$ $L_\odot$)} &
\colhead{($h_{70}^{-2}$ $L_\odot$)} &
\colhead{($h_{70}^{-2}$ $L_\odot$)} &
\colhead{($h_{70}^{-2}$ $L_\odot$)} & &
\colhead{Ref.} 
}
\startdata
A85\dotfill & 0.0521 & 6.10$^{0.12}_{0.12}$ & $6.62^{0.11}_{0.11}{\rm E}13$ & $2.41^{0.04}_{0.04}{\rm E}13$ & $0.564^{0.003}_{0.003}$ & $0.278^{0.001}_{0.001}$ & $2.80^{0.03}_{0.03}{\rm E}44$ & $2.31^{0.01}_{0.01}{\rm E}44$ & $1.38^{0.03}_{0.03}{\rm E}44$ & $5.99^{0.06}_{0.06}{\rm E}12$ & $\times$ & 9 \\ [.4em] 
A119\dotfill & 0.0444 & 5.80$^{0.36}_{0.36}$ & $4.93^{0.16}_{0.16}{\rm E}13$ & $1.42^{0.05}_{0.05}{\rm E}13$ & $0.471^{0.004}_{0.004}$ & \ldots & $1.03^{0.03}_{0.03}{\rm E}44$ & $6.59^{0.08}_{0.08}{\rm E}43$ & $8.16^{0.24}_{0.24}{\rm E}43$ & $6.46^{0.07}_{0.07}{\rm E}12$ & & 9 \\ [.4em] 
A262\dotfill & 0.0163 & 2.21$^{0.03}_{0.03}$ & $9.88^{0.31}_{0.31}{\rm E}12$ & $3.36^{0.10}_{0.10}{\rm E}12$ & $0.189^{0.002}_{0.002}$ & \ldots & $2.54^{0.09}_{0.09}{\rm E}43$ & $2.00^{0.03}_{0.03}{\rm E}43$ & $1.42^{0.09}_{0.09}{\rm E}43$ & $1.99^{0.07}_{0.07}{\rm E}12$ & $\times$ & 13 \\ [.4em] 
A401\dotfill & 0.0748 & 8.30$^{0.31}_{0.31}$ & $1.11^{0.03}_{0.03}{\rm E}14$ & $3.84^{0.10}_{0.10}{\rm E}13$ & $0.726^{0.005}_{0.005}$ & $0.355^{0.001}_{0.001}$ & $3.85^{0.05}_{0.05}{\rm E}44$ & $2.99^{0.02}_{0.02}{\rm E}44$ & $2.30^{0.03}_{0.03}{\rm E}44$ & \ldots & & 9 \\ [.4em] 
A426\dotfill & 0.0179 & 5.28$^{0.03}_{0.03}$ & $7.11^{0.16}_{0.16}{\rm E}13$ & $2.68^{0.06}_{0.06}{\rm E}13$ & $0.596^{0.009}_{0.009}$ & $0.319^{0.001}_{0.001}$ & \ldots & $4.05^{0.02}_{0.02}{\rm E}44$ & \ldots & $7.92^{0.06}_{0.06}{\rm E}12$ & $\times$ & 13 \\ [.4em] 
A478\dotfill & 0.0882 & 6.84$^{0.13}_{0.13}$ & $8.43^{0.17}_{0.17}{\rm E}13$ & $3.50^{0.07}_{0.07}{\rm E}13$ & $0.621^{0.003}_{0.003}$ & $0.351^{0.000}_{0.000}$ & $5.29^{0.04}_{0.04}{\rm E}44$ & $4.74^{0.03}_{0.03}{\rm E}44$ & $1.98^{0.02}_{0.02}{\rm E}44$ & $7.96^{0.09}_{0.09}{\rm E}12$ & $\times$ & 7 \\ [.4em] 
A496\dotfill & 0.0331 & 3.91$^{0.04}_{0.04}$ & $2.96^{0.05}_{0.05}{\rm E}13$ & $1.11^{0.02}_{0.02}{\rm E}13$ & $0.375^{0.004}_{0.004}$ & $0.175^{0.000}_{0.000}$ & $1.22^{0.02}_{0.02}{\rm E}44$ & $1.04^{0.01}_{0.01}{\rm E}44$ & $5.57^{0.21}_{0.21}{\rm E}43$ & $3.91^{0.05}_{0.05}{\rm E}12$ & $\times$ & 2 \\ [.4em] 
A644\dotfill & 0.0711 & 6.59$^{0.10}_{0.10}$ & $6.54^{0.20}_{0.20}{\rm E}13$ & $2.61^{0.08}_{0.08}{\rm E}13$ & $0.533^{0.002}_{0.002}$ & $0.295^{0.000}_{0.000}$ & $2.72^{0.03}_{0.03}{\rm E}44$ & $2.36^{0.02}_{0.02}{\rm E}44$ & $1.25^{0.02}_{0.02}{\rm E}44$ & $6.34^{0.06}_{0.06}{\rm E}12$ & $\times$ & 2 \\ [.4em] 
A754\dotfill & 0.0542 & 8.50$^{0.30}_{0.30}$ & $1.04^{0.11}_{0.11}{\rm E}14$ & $3.32^{0.34}_{0.34}{\rm E}13$ & $0.665^{0.003}_{0.003}$ & $0.349^{0.001}_{0.001}$ & $2.83^{0.04}_{0.04}{\rm E}44$ & $2.04^{0.01}_{0.01}{\rm E}44$ & $1.99^{0.03}_{0.03}{\rm E}44$ & $9.28^{0.05}_{0.05}{\rm E}12$ & & 5 \\ [.4em] 
A780\dotfill & 0.0565 & 3.80$^{0.12}_{0.12}$ & $3.26^{0.05}_{0.05}{\rm E}13$ & $1.39^{0.02}_{0.02}{\rm E}13$ & $0.408^{0.001}_{0.001}$ & $0.233^{0.001}_{0.001}$ & $2.08^{0.02}_{0.02}{\rm E}44$ & $1.87^{0.01}_{0.01}{\rm E}44$ & $7.86^{0.14}_{0.14}{\rm E}43$ & $3.09^{0.02}_{0.02}{\rm E}12$ & $\times$ & 9 \\ [.4em] 
A1060\dotfill & 0.0124 & 3.10$^{0.12}_{0.12}$ & $9.96^{0.37}_{0.37}{\rm E}12$ & 
$4.12^{0.15}_{0.15}{\rm E}12$ & $0.186^{0.002}_{0.002}$ & \ldots & \ldots & $1.89^{0.06}_{0.06}{\rm E}43$ & \ldots & $2.39^{0.06}_{0.06}{\rm E}12$ & $\times$ & 12 \\ [.4em] 
A1367\dotfill & 0.0214 & 3.50$^{0.11}_{0.11}$ & $2.19^{0.05}_{0.05}{\rm E}13$ & $6.13^{0.14}_{0.14}{\rm E}12$ & $0.312^{0.002}_{0.002}$ & \ldots & $4.48^{0.15}_{0.15}{\rm E}43$ & $2.80^{0.04}_{0.04}{\rm E}43$ & $3.59^{0.14}_{0.14}{\rm E}43$ & $3.81^{0.05}_{0.05}{\rm E}12$ & & 2 \\ [.4em] 
A1651\dotfill & 0.0846 & 6.30$^{0.30}_{0.30}$ & $6.39^{0.17}_{0.17}{\rm E}13$ & $2.40^{0.06}_{0.06}{\rm E}13$ & $0.522^{0.003}_{0.003}$ & $0.288^{0.000}_{0.000}$ & $2.46^{0.05}_{0.05}{\rm E}44$ & $2.09^{0.03}_{0.03}{\rm E}44$ & $1.25^{0.03}_{0.03}{\rm E}44$ & $7.82^{0.05}_{0.05}{\rm E}12$ & $\times$ & 9 \\ [.4em] 
A1656\dotfill & 0.0231 & 8.21$^{0.10}_{0.10}$ & $8.74^{0.40}_{0.40}{\rm E}13$ & $3.12^{0.14}_{0.14}{\rm E}13$ & $0.610^{0.006}_{0.006}$ & $0.286^{0.001}_{0.001}$ & $2.30^{0.06}_{0.06}{\rm E}44$ & $1.87^{0.01}_{0.01}{\rm E}44$ & $1.41^{0.06}_{0.06}{\rm E}44$ & $8.94^{0.05}_{0.05}{\rm E}12$ & & 6 \\ [.4em] 
A1689\dotfill & 0.1840 & 9.23$^{0.17}_{0.17}$ & $1.23^{0.03}_{0.03}{\rm E}14$ & $5.08^{0.14}_{0.14}{\rm E}13$ & $0.704^{0.006}_{0.006}$ & $0.425^{0.000}_{0.000}$ & $6.65^{0.09}_{0.09}{\rm E}44$ & $6.04^{0.07}_{0.07}{\rm E}44$ & $2.90^{0.05}_{0.05}{\rm E}44$ & \ldots & $\times$ & 11 \\ [.4em] 
A1795\dotfill & 0.0622 & 5.34$^{0.07}_{0.07}$ & $5.57^{0.07}_{0.07}{\rm E}13$ & $2.32^{0.03}_{0.03}{\rm E}13$ & $0.521^{0.003}_{0.003}$ & $0.286^{0.001}_{0.001}$ & $3.35^{0.03}_{0.03}{\rm E}44$ & $3.00^{0.01}_{0.01}{\rm E}44$ & $1.26^{0.02}_{0.02}{\rm E}44$ & $4.80^{0.04}_{0.04}{\rm E}12$ & $\times$ & 2 \\ [.4em] 
A2029\dotfill & 0.0766 & 8.70$^{0.18}_{0.18}$ & $1.05^{0.02}_{0.02}{\rm E}14$ & $4.24^{0.07}_{0.07}{\rm E}13$ & $0.653^{0.008}_{0.008}$ & $0.351^{0.001}_{0.001}$ & $5.57^{0.09}_{0.09}{\rm E}44$ & $5.03^{0.03}_{0.03}{\rm E}44$ & $1.90^{0.08}_{0.08}{\rm E}44$ & $8.12^{0.06}_{0.06}{\rm E}12$ & $\times$ & 9 \\ [.4em] 
A2052\dotfill & 0.0353 & 3.03$^{0.04}_{0.04}$ & $1.81^{0.04}_{0.04}{\rm E}13$ & $7.38^{0.16}_{0.16}{\rm E}12$ & $0.297^{0.003}_{0.003}$ & $0.148^{0.000}_{0.000}$ & $8.13^{0.26}_{0.26}{\rm E}43$ & $7.18^{0.09}_{0.09}{\rm E}43$ & $3.30^{0.24}_{0.24}{\rm E}43$ & $3.19^{0.05}_{0.05}{\rm E}12$ & $\times$ & 13 \\ [.4em] 
A2063\dotfill & 0.0355 & 3.68$^{0.07}_{0.07}$ & $2.33^{0.05}_{0.05}{\rm E}13$ & $8.79^{0.20}_{0.20}{\rm E}12$ & $0.313^{0.004}_{0.004}$ & $0.141^{0.001}_{0.001}$ & $6.81^{0.35}_{0.35}{\rm E}43$ & $5.63^{0.10}_{0.10}{\rm E}43$ & $3.72^{0.33}_{0.33}{\rm E}43$ & $3.34^{0.04}_{0.04}{\rm E}12$ & $\times$ & 4 \\ [.4em] 
A2142\dotfill & 0.0894 & 8.68$^{0.12}_{0.12}$ & $1.38^{0.01}_{0.01}{\rm E}14$ & $5.04^{0.04}_{0.04}{\rm E}13$ & $0.787^{0.004}_{0.004}$ & $0.430^{0.001}_{0.001}$ & $6.65^{0.05}_{0.05}{\rm E}44$ & $5.52^{0.03}_{0.03}{\rm E}44$ & $3.28^{0.04}_{0.04}{\rm E}44$ & $7.20^{0.09}_{0.09}{\rm E}12$ & $\times$ & 2 \\ [.4em] 
A2199\dotfill & 0.0299 & 4.10$^{0.05}_{0.05}$ & $2.94^{0.02}_{0.02}{\rm E}13$ & $1.19^{0.01}_{0.01}{\rm E}13$ & $0.356^{0.002}_{0.002}$ & $0.187^{0.000}_{0.000}$ & $1.28^{0.01}_{0.01}{\rm E}44$ & $1.13^{0.00}_{0.00}{\rm E}44$ & $5.03^{0.14}_{0.14}{\rm E}43$ & $4.25^{0.04}_{0.04}{\rm E}12$ & $\times$ & 4 \\ [.4em] 
A2204\dotfill & 0.1524 & 7.21$^{0.25}_{0.25}$ & $1.16^{0.06}_{0.06}{\rm E}14$ & $4.29^{0.22}_{0.22}{\rm E}13$ & $0.723^{0.013}_{0.013}$ & $0.400^{0.002}_{0.002}$ & $8.45^{0.16}_{0.16}{\rm E}44$ & $7.55^{0.11}_{0.11}{\rm E}44$ & $3.08^{0.10}_{0.10}{\rm E}44$ & \ldots & $\times$ & 13 \\ [.4em] 
A2244\dotfill & 0.0970 & 7.10$^{2.40}_{1.50}$ & $6.89^{0.98}_{0.98}{\rm E}13$ & $2.60^{0.37}_{0.37}{\rm E}13$ & $0.529^{0.002}_{0.002}$ & $0.275^{0.001}_{0.001}$ & $2.65^{0.07}_{0.07}{\rm E}44$ & $2.24^{0.05}_{0.05}{\rm E}44$ & $1.19^{0.03}_{0.03}{\rm E}44$ & \ldots & $\times$ & 2 \\ [.4em] 
A2255\dotfill & 0.0809 & 6.87$^{0.20}_{0.20}$ & $7.17^{0.14}_{0.14}{\rm E}13$ & $2.23^{0.04}_{0.04}{\rm E}13$ & $0.598^{0.003}_{0.003}$ & \ldots & $1.65^{0.03}_{0.03}{\rm E}44$ & $1.13^{0.01}_{0.01}{\rm E}44$ & $1.27^{0.02}_{0.02}{\rm E}44$ & $8.70^{0.08}_{0.08}{\rm E}12$ & & 13 \\ [.4em] 
A2256\dotfill & 0.0581 & 7.51$^{0.11}_{0.11}$ & $8.70^{0.20}_{0.20}{\rm E}13$ & $3.20^{0.07}_{0.07}{\rm E}13$ & $0.639^{0.003}_{0.003}$ & $0.323^{0.000}_{0.000}$ & $2.69^{0.04}_{0.04}{\rm E}44$ & $2.14^{0.01}_{0.01}{\rm E}44$ & $1.77^{0.03}_{0.03}{\rm E}44$ & $1.01^{0.00}_{0.00}{\rm E}13$ & & 2 \\ [.4em] 
A2319\dotfill & 0.0555 & 9.12$^{0.09}_{0.09}$ & $1.40^{0.04}_{0.04}{\rm E}14$ & $4.45^{0.13}_{0.13}{\rm E}13$ & $0.833^{0.010}_{0.010}$ & $0.393^{0.001}_{0.001}$ & $4.84^{0.09}_{0.09}{\rm E}44$ & $3.66^{0.04}_{0.04}{\rm E}44$ & $3.09^{0.08}_{0.08}{\rm E}44$ & $1.28^{0.00}_{0.00}{\rm E}13$ & & 1 \\ [.4em] 
A2597\dotfill & 0.0852 & 3.60$^{0.12}_{0.12}$ & $2.79^{0.15}_{0.15}{\rm E}13$ & $1.17^{0.06}_{0.06}{\rm E}13$ & $0.386^{0.002}_{0.002}$ & $0.217^{0.000}_{0.000}$ & $2.19^{0.03}_{0.03}{\rm E}44$ & $2.01^{0.03}_{0.03}{\rm E}44$ & $6.85^{0.13}_{0.13}{\rm E}43$ & \ldots & $\times$ & 9 \\ [.4em] 
A3112\dotfill & 0.0703 & 4.70$^{0.24}_{0.24}$ & $4.05^{0.24}_{0.24}{\rm E}13$ & $1.53^{0.09}_{0.09}{\rm E}13$ & $0.417^{0.002}_{0.002}$ & $0.225^{0.000}_{0.000}$ & $2.17^{0.03}_{0.03}{\rm E}44$ & $1.91^{0.02}_{0.02}{\rm E}44$ & $7.86^{0.21}_{0.21}{\rm E}43$ & $4.11^{0.08}_{0.08}{\rm E}12$ & $\times$ & 9 \\ [.4em] 
A3158\dotfill & 0.0590 & 5.77$^{0.10}_{0.05}$ & $5.49^{0.12}_{0.12}{\rm E}13$ & $1.97^{0.04}_{0.04}{\rm E}13$ & $0.513^{0.004}_{0.004}$ & $0.255^{0.001}_{0.001}$ & $1.77^{0.03}_{0.03}{\rm E}44$ & $1.39^{0.02}_{0.02}{\rm E}44$ & $1.11^{0.02}_{0.02}{\rm E}44$ & $6.89^{0.05}_{0.05}{\rm E}12$ & & 13 \\ [.4em] 
A3266\dotfill & 0.0545 & 7.70$^{0.48}_{0.48}$ & $8.41^{0.30}_{0.30}{\rm E}13$ & $2.79^{0.10}_{0.10}{\rm E}13$ & $0.663^{0.003}_{0.003}$ & $0.273^{0.001}_{0.001}$ & $2.43^{0.04}_{0.04}{\rm E}44$ & $1.83^{0.01}_{0.01}{\rm E}44$ & $1.71^{0.04}_{0.04}{\rm E}44$ & $8.38^{0.06}_{0.06}{\rm E}12$ & & 9 \\ [.4em] 
A3391\dotfill & 0.0550 & 5.70$^{0.42}_{0.42}$ & $4.21^{0.21}_{0.21}{\rm E}13$ & $1.31^{0.07}_{0.07}{\rm E}13$ & $0.403^{0.003}_{0.003}$ & \ldots & $8.99^{0.35}_{0.35}{\rm E}43$ & $6.34^{0.13}_{0.13}{\rm E}43$ & $6.22^{0.29}_{0.29}{\rm E}43$ & $5.84^{0.09}_{0.09}{\rm E}12$ & & 9 \\ [.4em] 
A3526\dotfill & 0.0101 & 3.54$^{0.08}_{0.08}$ & $1.52^{0.03}_{0.03}{\rm E}13$ & $5.39^{0.09}_{0.09}{\rm E}12$ & $0.227^{0.003}_{0.003}$ & \ldots & \ldots & $3.22^{0.07}_{0.07}{\rm E}43$ & \ldots & $3.66^{0.16}_{0.16}{\rm E}12$ & $\times$ & 1 \\ [.4em] 
A3532\dotfill & 0.0553 & 4.58$^{0.12}_{0.10}$ & $3.23^{0.09}_{0.09}{\rm E}13$ & $1.04^{0.03}_{0.03}{\rm E}13$ & $0.380^{0.005}_{0.005}$ & \ldots & $7.66^{0.26}_{0.26}{\rm E}43$ & $5.49^{0.10}_{0.10}{\rm E}43$ & $5.40^{0.22}_{0.22}{\rm E}43$ & \ldots & & 11 \\ [.4em] 
A3558\dotfill & 0.0477 & 5.70$^{0.12}_{0.12}$ & $6.32^{0.10}_{0.10}{\rm E}13$ & $2.04^{0.03}_{0.03}{\rm E}13$ & $0.575^{0.007}_{0.007}$ & $0.261^{0.001}_{0.001}$ & $2.16^{0.04}_{0.04}{\rm E}44$ & $1.59^{0.01}_{0.01}{\rm E}44$ & $1.43^{0.04}_{0.04}{\rm E}44$ & $1.11^{0.01}_{0.01}{\rm E}13$ & & 3 \\ [.4em] 
A3562\dotfill & 0.0502 & 5.16$^{0.16}_{0.16}$ & $3.66^{0.10}_{0.10}{\rm E}13$ & $1.11^{0.03}_{0.03}{\rm E}13$ & $0.385^{0.005}_{0.005}$ & $0.141^{0.000}_{0.000}$ & $9.97^{0.32}_{0.32}{\rm E}43$ & $6.77^{0.09}_{0.09}{\rm E}43$ & $6.58^{0.29}_{0.29}{\rm E}43$ & $3.56^{0.09}_{0.09}{\rm E}12$ & $\times$ & 13 \\ [.4em] 
A3571\dotfill & 0.0397 & 6.90$^{0.18}_{0.18}$ & $7.19^{0.21}_{0.21}{\rm E}13$ & $2.66^{0.08}_{0.08}{\rm E}13$ & $0.549^{0.005}_{0.005}$ & $0.287^{0.001}_{0.001}$ & $2.60^{0.05}_{0.05}{\rm E}44$ & $2.20^{0.02}_{0.02}{\rm E}44$ & $1.29^{0.05}_{0.05}{\rm E}44$ & \ldots & $\times$ & 9 \\ [.4em] 
A3667\dotfill & 0.0530 & 7.00$^{0.36}_{0.36}$ & $8.57^{0.23}_{0.23}{\rm E}13$ & $2.62^{0.07}_{0.07}{\rm E}13$ & $0.720^{0.007}_{0.007}$ & $0.283^{0.001}_{0.001}$ & $2.66^{0.06}_{0.06}{\rm E}44$ & $1.78^{0.02}_{0.02}{\rm E}44$ & $1.88^{0.05}_{0.05}{\rm E}44$ & $8.65^{0.06}_{0.06}{\rm E}12$ & & 9 \\ [.4em] 
A4038\dotfill & 0.0281 & 3.15$^{0.03}_{0.03}$ & $1.84^{0.11}_{0.11}{\rm E}13$ & $6.58^{0.40}_{0.40}{\rm E}12$ & $0.271^{0.001}_{0.001}$ & $0.126^{0.000}_{0.000}$ & $6.13^{0.15}_{0.15}{\rm E}43$ & $5.24^{0.07}_{0.07}{\rm E}43$ & $2.72^{0.12}_{0.12}{\rm E}43$ & $2.85^{0.06}_{0.06}{\rm E}12$ & $\times$ & 13 \\ [.4em] 
A4059\dotfill & 0.0456 & 4.10$^{0.18}_{0.18}$ & $2.69^{0.08}_{0.08}{\rm E}13$ & $9.80^{0.30}_{0.30}{\rm E}12$ & $0.335^{0.001}_{0.001}$ & $0.160^{0.000}_{0.000}$ & $9.02^{0.20}_{0.20}{\rm E}43$ & $7.77^{0.10}_{0.10}{\rm E}43$ & $4.07^{0.15}_{0.15}{\rm E}43$ & $3.12^{0.10}_{0.10}{\rm E}12$ & $\times$ & 9 \\ [.4em] 
0745-19\dotfill & 0.1028 & 7.21$^{0.11}_{0.11}$ & $1.13^{0.02}_{0.02}{\rm E}14$ & $4.44^{0.09}_{0.09}{\rm E}13$ & $0.708^{0.006}_{0.006}$ & $0.396^{0.001}_{0.001}$ & $9.28^{0.11}_{0.11}{\rm E}44$ & $8.44^{0.09}_{0.09}{\rm E}44$ & $2.86^{0.05}_{0.05}{\rm E}44$ & \ldots & $\times$ & 13 \\ [.4em] 
AWM7\dotfill & 0.0172 & 3.90$^{0.12}_{0.12}$ & $2.26^{0.08}_{0.08}{\rm E}13$ & $8.80^{0.32}_{0.32}{\rm E}12$ & $0.314^{0.002}_{0.002}$ & $0.141^{0.000}_{0.000}$ & $6.56^{0.21}_{0.21}{\rm E}43$ & $5.77^{0.06}_{0.06}{\rm E}43$ & $3.18^{0.20}_{0.20}{\rm E}43$ & $2.90^{0.06}_{0.06}{\rm E}12$ & $\times$ & 8 \\ [.4em] 
Cygnus A\dotfill & 0.0570 & 6.50$^{0.36}_{0.36}$ & $7.54^{0.36}_{0.36}{\rm E}13$ & $2.39^{0.12}_{0.12}{\rm E}13$ & $0.657^{0.015}_{0.015}$ & $0.249^{0.002}_{0.002}$ & $3.53^{0.15}_{0.15}{\rm E}44$ & $2.69^{0.04}_{0.04}{\rm E}44$ & $1.64^{0.14}_{0.14}{\rm E}44$ & \ldots & $\times$ & 9 \\ [.4em] 
MKW3S\dotfill & 0.0430 & 3.50$^{0.12}_{0.12}$ & $2.14^{0.15}_{0.15}{\rm E}13$ & $8.03^{0.57}_{0.57}{\rm E}12$ & $0.295^{0.004}_{0.004}$ & $0.155^{0.000}_{0.000}$ & $8.26^{0.32}_{0.32}{\rm E}43$ & $7.45^{0.10}_{0.10}{\rm E}43$ & $3.02^{0.30}_{0.30}{\rm E}43$ & $1.96^{0.08}_{0.08}{\rm E}12$ & $\times$ & 9 \\ [.4em] 
Ophiuchus\dotfill & 0.0280 & 9.80$^{0.61}_{0.61}$ & $1.04^{0.05}_{0.05}{\rm E}14$ & $4.26^{0.21}_{0.21}{\rm E}13$ & $0.610^{0.011}_{0.011}$ & $0.339^{0.001}_{0.001}$ & \ldots & $3.40^{0.04}_{0.04}{\rm E}44$ & \ldots & \ldots & $\times$ & 10 \\ [.4em] 
Tria Aust\dotfill & 0.0510 & 9.50$^{0.42}_{0.42}$ & $1.26^{0.03}_{0.03}{\rm E}14$ & $4.53^{0.12}_{0.12}{\rm E}13$ & $0.730^{0.011}_{0.011}$ & $0.370^{0.001}_{0.001}$ & $3.97^{0.08}_{0.08}{\rm E}44$ & $3.19^{0.03}_{0.03}{\rm E}44$ & $2.36^{0.07}_{0.07}{\rm E}44$ & $1.20^{0.01}_{0.01}{\rm E}13$ & & 9 
\enddata
\label{tab:clusterinfo}

\tablerefs{
(1) \citealt{arnaud99}
(2) \citealt{david93}
(3) \citealt{day91}
(4) \citealt{fukazawa98}
(5) \citealt{henriksen96}
(6) \citealt{hughes93}
(7) \citealt{johnstone92}
(8) \citealt{markevitch97}
(9) \citealt{markevitch98}
(10) \citealt{matsuzawa96}
(11) \citealt{reiprich02}
(12) \citealt{tamura96}
(13) \citealt{dwhite00}
}
\end{deluxetable*}
\tabletypesize{\footnotesize}

We present here a joint analysis of a flux-limited sample of 45 nearby clusters observed with the \ROSAT\ PSPC and the Two Micron All Sky Survey (2MASS), and an ensemble of 45 simulated clusters evolved in a cosmological context. We begin with the scaling relations from the observational sample in \S~\ref{sec:relations}. In \S~\ref{sec:cores} we discuss the effects of cool cores in our sample. We discuss a method of correcting for cool core effects so that we can examine only structural, merger-related scatter. We discuss the use of peak surface brightness as an indication of cool core strength in \S~\ref{sec:bright}. \S~\ref{sec:subscatter} reports the study of the relationship between displacement from scaling relations and substructure in both observed and simulated clusters. Finally, we list our conclusions in \S~\ref{sec:concl}.
 
Throughout the paper we assume a $\Lambda$CDM cosmology with $\Omega_M=0.3$ and 
$\Omega_\Lambda = 0.7$, and use a Hubble parameter of
$H_0 = 70\,h_{70}$~km~s$^{-1}$~Mpc$^{-1}$. All uncertainties are 68\% confidence, or 1 $\sigma$.


\section{Observed Scaling Relations}
\label{sec:relations}

We study an ensemble of 45 members of the \citet{edge90} flux-limited sample, observed with \ROSAT\ PSPC. We use the same reduced imaging data as \citet{mohr99} (hereafter MME). These data have a pixel scale of $14\arcsec.947$, and an energy range of 0.5--2.0 keV. The resolution of PSPC is, of course, poorer than the current generation of X-ray instruments, but none of the observables we are measuring require higher resolution. For details of the reduction, see MME.
For 34 of these clusters we also use measurements of $K$-band light from \citet{lin04}; the reduction, measurements, and uncertainties are discussed therein.

We use previously published emission-weighted mean temperatures measured with {\it Ginga} and {\it ASCA}, except for A2244, where we use a temperature measured with the {\it Einstein} MPC, and A3532, for which no appropriate temperature with uncertainties can be found in the literature, and for which we use a temperature derived from a luminosity--temperature relation \citep{reiprich02}. We use PSF-corrected, cluster X-ray peak surface brightness values from MME, but all other X-ray observable quantities are measured as part of this analysis.
We divide the sample into cool core (CC) and non-cool core (NCC) clusters based on published central cooling times \citep{peres98}. We adopt the classification of \citet{mohr97a} in which CC clusters are those with central cooling times at least 3$\sigma$ below 7.1 Gyr (for $H_0 = 70$~km~s$^{-1}$~Mpc$^{-1}$; 10 Gyr for $H_0 = 50$~km~s$^{-1}$~Mpc$^{-1}$).
By this measure our sample of 45 clusters contains 30 CC and 15 NCC clusters. 
Basic information about the sample, plus the measured observables, is given in Table~\ref{tab:clusterinfo}.

In this section we first examine the X-ray luminosity--temperature relation in detail as an example of the scaling relations we are using.  We point out features that will be common to all relations and demonstrate that the source of scatter is true structural variation among clusters, not measurement uncertainties. We then present the remaining X-ray scaling relations and quantify their scatter as well. Finally, we present the near-IR luminosity--temperature relation and discuss the additional information available from the galaxy properties.

\subsection{X-ray Luminosity--Temperature Relation}

We begin by examining the scaling relation constructed from \LXfive\ (i.e., the X-ray luminosity projected within a distance corresponding to the virial radius \rfive) and the emission-weighted mean temperature \Tx\ (all temperature values given in this paper are emission-weighted X-ray temperatures). The luminosity is measured in the 0.5--2 keV band. The virial radius  \rd is the radius within which the mean density is $\Delta$ times the critical density of the universe; these radii are obtained from observed $M_{\rm vir}$--\Tx\ virial scaling relations. For \rfive\ we use a mass--temperature relation obtained using clusters with masses greater than $3.6\times10^{13} M_\odot$ \citep{finoguenov01}, which gives

\begin{equation}
r_{500} = 0.447  ~ h_{70}^{-1} ~ T_X^{0.527} ~ {\rm Mpc} ~ .
\end{equation} 

Note that in our analysis we use the emission-weighted mean temperature rather than cool core-corrected emission-weighted mean temperatures.  The \citet{finoguenov01} relation is derived using temperature profiles measured with ASCA, but their emission-weighted mean temperatures have been corrected for cool core effects.

One CC cluster in our sample, A3526, has a value of \rfive\ large enough that it extends beyond the edges of the PSPC image. Three other CC clusters (Ophiuchus, A426, and A1060) have values of \rfive\ that become too large for the images when the temperature is scaled to account for cool core effects in the analysis presented in \S~\ref{sec:cores}. We therefore exclude these four CC clusters from our analysis involving  \LXfive\ (and, below, from the analysis of the core-subtracted luminosity within \rfive).

We consider X-ray luminosity uncertainties from three sources. First, we measure luminosity projected within a virial radius determined by the temperature, so there is an uncertainty in the radius due to the uncertainty $\Delta T_{\rm X}$ in the measured cluster temperature \Tx. We thus measure the luminosity projected within radii determined from temperatures $T_{\rm X} + \Delta T_{\rm X}$ and $T_{\rm X} - \Delta T_{\rm X}$, and average the absolute difference between these luminosities and the luminosity measured at the virial radius. There is also some uncertainty in the X-ray images; we use error images created as described in MME, and sum the error within the region of interest just as we do for the X-ray image counts. Finally, there is an uncertainty in the background, which we account for by raising and lowering the background level by 10\%, measuring the luminosity at each level, and averaging the deviations from the value measured with the standard background level. The temperature-induced uncertainty, X-ray image uncertainty, and background uncertainty are added in quadrature to obtain the total uncertainty in the X-ray luminosity.

 \begin{inlinefigure}
   {
   \begin{center}
   \epsfxsize=7.cm
   \begin{minipage}{\epsfxsize}\epsffile{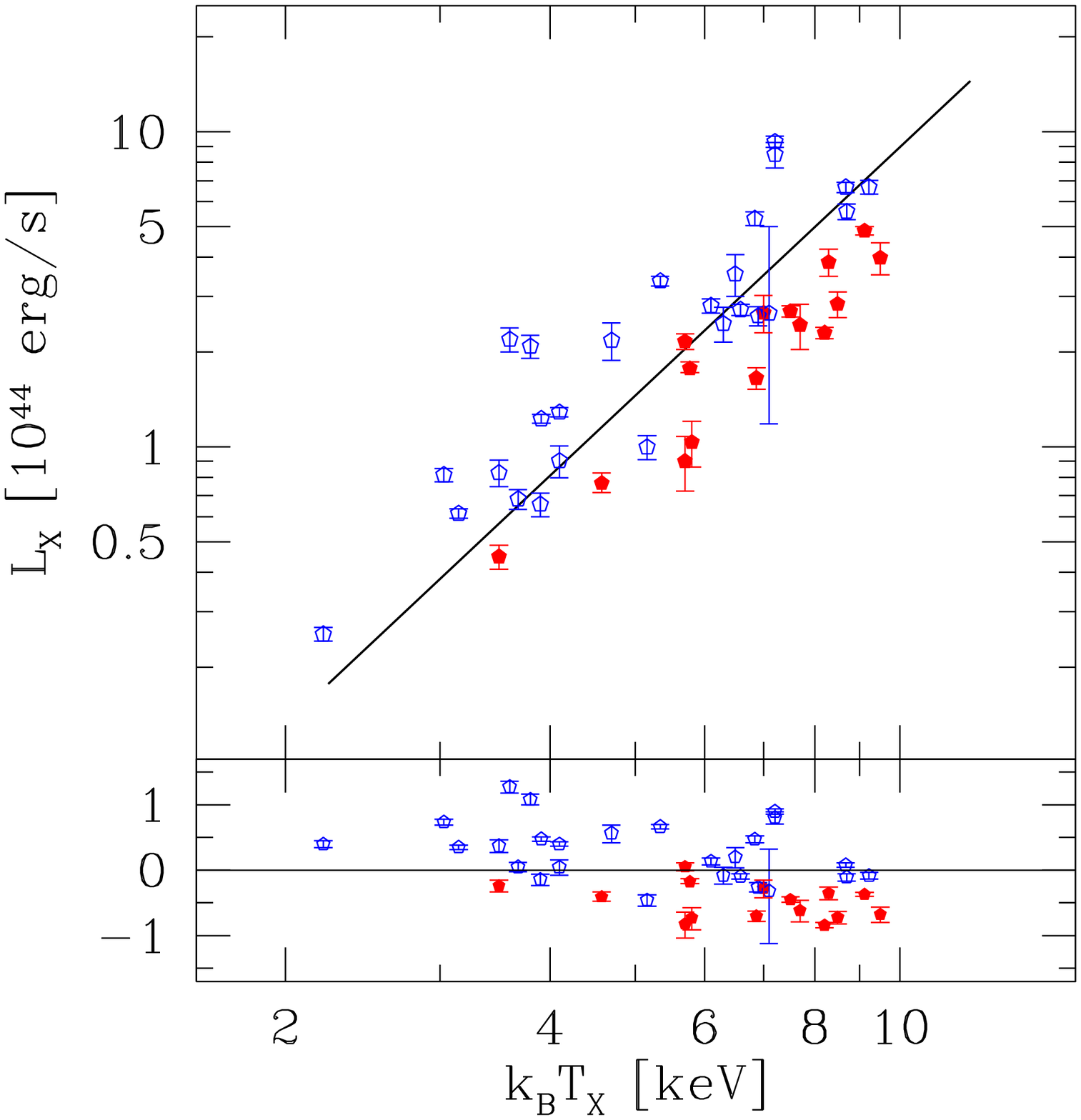}\end{minipage}
   \end{center}}
   \figcaption{\label{fig:LXerr} 
   X-ray luminosity projected within \rfive\ ({\it top}), and deviation of clusters from the best fit relation ({\it bottom}). Open and filled markers correspond to CC and NCC clusters, respectively. The uncertainties include both the measurement uncertainty in the luminosity and an effective temperature contribution to the luminosity uncertainty as described in the text.
     }
\end{inlinefigure}

When we examine the deviation of a cluster from a scaling relation, we do so in one dimension only, e.g., only for luminosity in the luminosity--temperature relation. We thus include a temperature component to the uncertainty in each observable, which we approximate using the best fit power law to the observed scaling relation. That is, for each observable $\mathcal{O}$ we determine the scaling relation exponent $\alpha$ (i.e., $\mathcal{O} \propto T_{\rm X}^\alpha$), and find a fractional uncertainty due to the temperature,

\begin{equation}
\frac{\sigma_{\mathcal{O}}}{\mathcal{O}} = \alpha \frac{\sigma_{T_{\rm X}}}{T_{\rm X}} ~,
\end{equation}

\noindent which is added in quadrature to the measurement uncertainty.

Figure~\ref{fig:LXerr} shows the \LXfive--\Tx\ scaling relation, including uncertainties. One immediately notices that there is a separation of the CC and NCC populations on the relation, i.e., the CC clusters tend to lie above the best-fit line, and the NCC clusters lie below; this relationship has been observed before \citep[e.g.,][]{fabian94c,markevitch98b,mccarthy04}. 
One part of the separation is likely a simple temperature bias; that is, CC clusters have a cool central region that does not significantly affect the structure of the cluster outside that region, but whose relatively high density leads to a significant bias toward lower emission-weighted mean temperatures relative to clusters without a cool core. The remainder of the separation is due primarily to the higher central gas density in CC clusters, which leads to greater X-ray luminosity at a given temperature.

\begin{deluxetable*}{ccccccccccc}
\tablewidth{0pt}
\tablecaption{Raw and Intrinsic Scatter in Scaling Relations}
\tablehead{
& \multicolumn{8}{c}{Observations} & & \multicolumn{1}{c}{Simulations\tablenotemark{a}} \cr
\cline{2-9} 
& \multicolumn{2}{c}{Original} & & \multicolumn{2}{c}{With CC}  && \multicolumn{2}{c}{3 Parameter} \cr
& \multicolumn{2}{c}{Relations} & & \multicolumn{2}{c}{Temp. Scaling} && \multicolumn{2}{c}{(\OTI)} \cr
\cline{2-3} \cline{5-6} \cline{8-9}
\colhead{Scaling Relation} & \colhead{\sraw} & \colhead{\sint} & & \colhead{\sraw} & \colhead{\sint} && \colhead{\sraw} & \colhead{\sint} & & \colhead{\sint} 
}
\startdata
\Micmfive--\Tx & 0.19 & 0.17 & & 0.18 & 0.17 && 0.16 & 0.15 & & 0.20  \\
\Micmtfive--\Tx & 0.23 & 0.22 & & 0.20 & 0.19 && 0.16 & 0.14  \\
\LXfive--\Tx & 0.53 & 0.53 & & 0.34 & 0.33 && 0.26 & 0.24 \\
\LXtfive--\Tx & 0.69 & 0.67 & & 0.39 & 0.39 && 0.36 & 0.34 \\
\LXCSfive--\Tx & 0.29 & 0.28 & & 0.25 & 0.23 && 0.23 & 0.21 & & 0.27  \\
\Rthree--\Tx & 0.14 & 0.14 & & 0.14 & 0.14 && 0.13 & 0.13 & & 0.10 \\
\Rone--\Tx & 0.17 & 0.17 & & 0.17 & 0.17 && 0.14 & 0.14 \\
\LNIRfive--\Tx & 0.19 & 0.19 & & 0.17 & 0.17 && 0.19 & 0.19 \\
\Rthree--\Micmfive & 0.06 & 0.06 \\
\Rthree--\LXCSfive & 0.06 & 0.05 \\
\Rthree--\LNIRfive & 0.16 & 0.16 
\enddata
\tablenotetext{a}{For simulations, \Micmfive\ is actual gas mass, not measured from mock observations, and the isophote for \RI\ is chosen to produce normalization similar to the observed \Rthree--\Tx\ relation.}
\label{tab:rawint}
\end{deluxetable*}

Because we will be examining the scatter about scaling relations and drawing conclusions about the effects of substructure on those relations, it is important to establish that the scatter is a real, intrinsic effect, and not due to measurement uncertainties. This is made plain qualitatively from the bottom portion of Figure~\ref{fig:LXerr}, which shows the deviation from the best-fit relation; the uncertainties in \LX\ are clearly smaller than the intrinsic scatter in the relation. To address the issue quantitatively, we measure both the raw (i.e., RMS) scatter and the intrinsic scatter, which we express in terms of $\ln \mathcal{O}$, by finding the values of $\sigma$ for which the reduced $\chi^2$ value for a given relation is unity. We find the intrinsic scatter \sint\ by adding a uniform value in quadrature to the measurement uncertainty for each cluster.

For the \LXfive--\Tx\ relation, we find \sint$=$\sraw$=0.53$ (these values are also listed in Table~\ref{tab:rawint}); that is, the intrinsic and raw scatter are the same to the precision given here, and so the intrinsic scatter is clearly much greater than the measurement uncertainties. We may thus be certain that the scatter about this relation is due to real structural differences between the clusters.

\vspace{-1pt}

\subsection{Other X-ray Scaling Relations}
\label{sec:measure}

In addition to the luminosity projected within \rfive, we measure \LXtfive, the luminosity projected within \rtfive; studying relations at different radii gives us information about the effects of structural variations on different scales within the cluster.   
To determine \rtfive\ we use an \M2500\--\Tx\ relation derived from \Chandra\ observations of relaxed intermediate-redshift clusters \citep{allen01}, which gives
\begin{equation}
r_{2500} = 0.227  ~ h_{70}^{-1} ~ T_X^{0.503} ~ {\rm Mpc} ~ .
\end{equation} 
We also measure the core-subtracted X-ray luminosity (\LXCS) within \rfive. We exclude the luminosity projected within a radius corresponding to $0.20 r_{500}$, which minimizes the effects of core structure 
\citep[e.g.,][]{neumann99}. This radius corresponds to $0.13 r_{200}$, where the virial radius \rtwo\ is calculated from \rfive\ by using an NFW dark matter density profile with concentration parameter $c=5$ \citep{navarro97}; the relationship is $r_{200} = 1.51r_{500}$.

\begin{inlinefigure}
   {
   \begin{center}
   \epsfxsize=6.cm
   \begin{minipage}{\epsfxsize}\epsffile{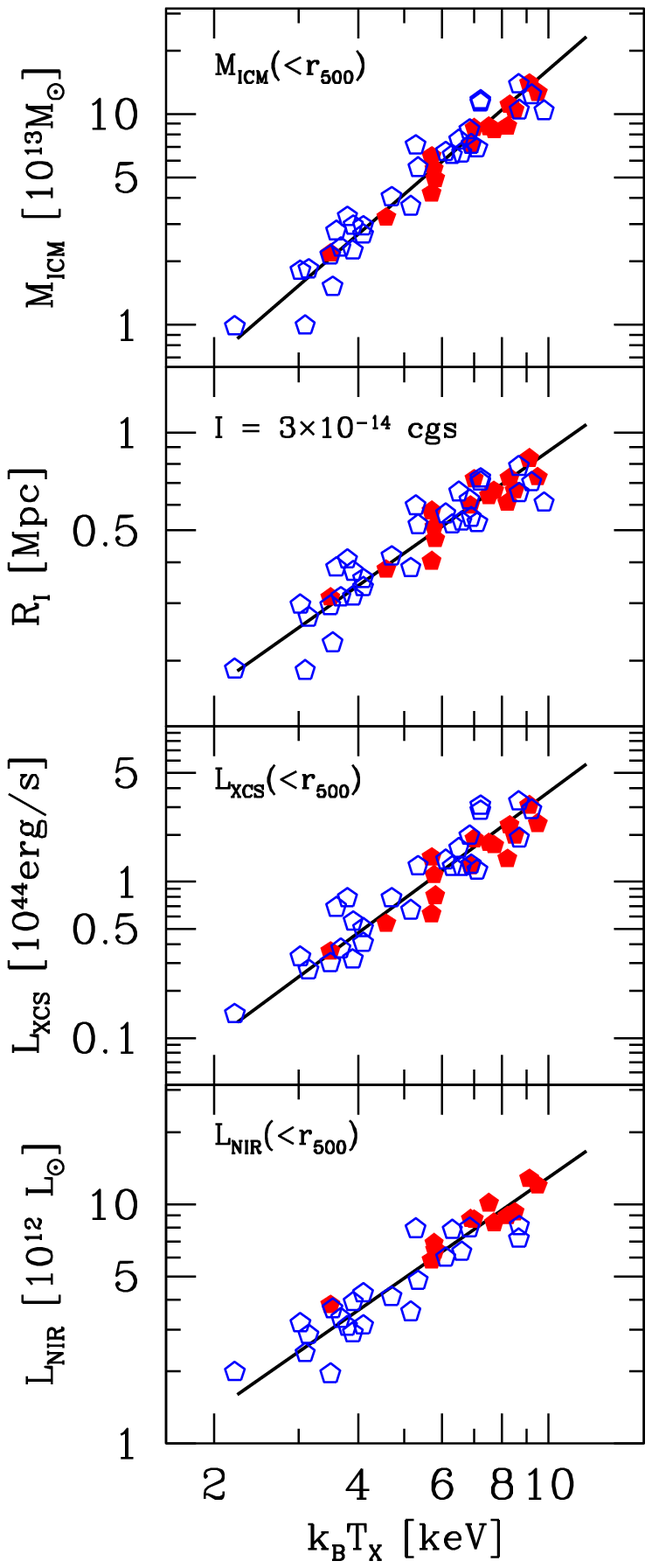}\end{minipage}
   \end{center}}


   \figcaption{\label{fig:scaling5} 
   From top to bottom, we show scaling relations for \Micm\ within \rfive, isophotal size for an isophote of  $3\times10^{-14}$ erg \ps\ cm$^{-2}$ arcmin$^{-2}$, the X--ray core subtracted luminosity projected within \rfive, and the $K$-band luminosity projected within \rfive, versus \Tx. The open and filled markers correspond to CC and NCC clusters, respectively.
     }
\end{inlinefigure}

We measure the ICM mass within \rfive\ and \rtfive. The masses are measured as discussed in MME, using the $\beta$-model parameters given in that paper (some clusters are fit with a double $\beta$-model), the emission-weighted mean temperature, and a measurement of the X-ray flux. We use the \Micmfive\ uncertainties from MME as a starting point, but we adjust the temperature uncertainty contribution to reflect the newer and more accurate temperatures available for some clusters. We adopt the same fractional uncertainty for \Micmtfive\ as for \Micmfive; the median fractional uncertainty is $\sim$3\%.

We determine the isophotal size \RI\ of a cluster by measuring the area \AI\ enclosed by a particular isophote {\it I} and finding the effective radius given by $A_I = \pi R_I^2$. We measure \RI\ for two isophotes: $3\times10^{-14}$ erg \ps\ cm$^{-2}$ arcmin$^{-2}$ and $1.5\times10^{-13}$ erg \ps\ cm$^{-2}$ arcmin$^{-2}$, in the 0.5--2 keV band. The lower isophote lies well outside the core region of the clusters, and so, like the luminosities and masses measured within \rfive, reflects cluster structure in a way largely unaffected by core substructure. The higher isophote is more reflective of core structure. For some clusters in our sample the central surface brightness does not rise (or barely rises) above the brighter isophote;  these eight clusters are excluded from all analysis at this isophote.

When measuring the isophotal size we include a background uncertainty of 10\%, calculate \RI\ with the higher and lower uncertainty, and take the average of the deviations from the standard background value to obtain the uncertainty in \RI . PSPC also has a $\sim$10\% uncertainty in its effective area. However, changing the effective area tends to simply shift the entire cluster population up or down in isophotal size, and does not affect the deviation of individual clusters from the relations; hence, we ignore this particular uncertainty.   In the soft X-ray band the conversion from PSPC counts s$^{-1}$ to physical flux is approximately independent of cluster temperature, and so temperature uncertainties do not lead to uncertainties in the measured isophotal size.

Each observable derives from a different integral over cluster structure:
\begin{displaymath}
\langle T_{\rm X} \rangle = \frac{1}{L_{\rm X}} \int n_e^2 \frac{\mu_e}{\mu_{\rm H}} \Lambda(T) T  d^3x 
\end{displaymath}

\begin{displaymath}
L_{\rm X} = \int n_e^2 \frac{\mu_e}{\mu_{\rm H}} \Lambda(T) d^3x
\end{displaymath}

\begin{displaymath}
M_{\rm ICM} = \int \mu_e m_p n_e d^3x ~ ,
\end{displaymath}

\noindent where $\Lambda(T)$ is the emissivity of the ICM gas, and the other symbols have their usual meanings. Isophotal size derives from cluster structure in a somewhat more complex way; see \citet{mohr00a} for a discussion. Thus by studying multiple observables, we are not simply increasing the size of our parameter space; we are, in fact, looking at several different ways of quantifying the structure of clusters.  By focusing on crude observables like these we are able to work in a regime where the measurement uncertainties are small compared to the intrinsic scatter.

The scaling relations for \Micmfive\, \Rthree\, and \LXCSfive\ are shown in Figure \ref{fig:scaling5}; in Figure \ref{fig:scaling25} we plot relations for \Micm\ within \rtfive, \LXtfive, and \Rone. For clarity we do not show measurement uncertainties; however, as shown for \LXfive\ in Figure~\ref{fig:LXerr} they are significantly smaller than the intrinsic scatter. The actual measurements of raw and intrinsic scatter are given for all relations in Table~\ref{tab:rawint}; the intrinsic scatter dominates the total scatter in all cases. 

\begin{inlinefigure}
   {
   \begin{center}
   \epsfxsize=7.cm
   \begin{minipage}{\epsfxsize}\epsffile{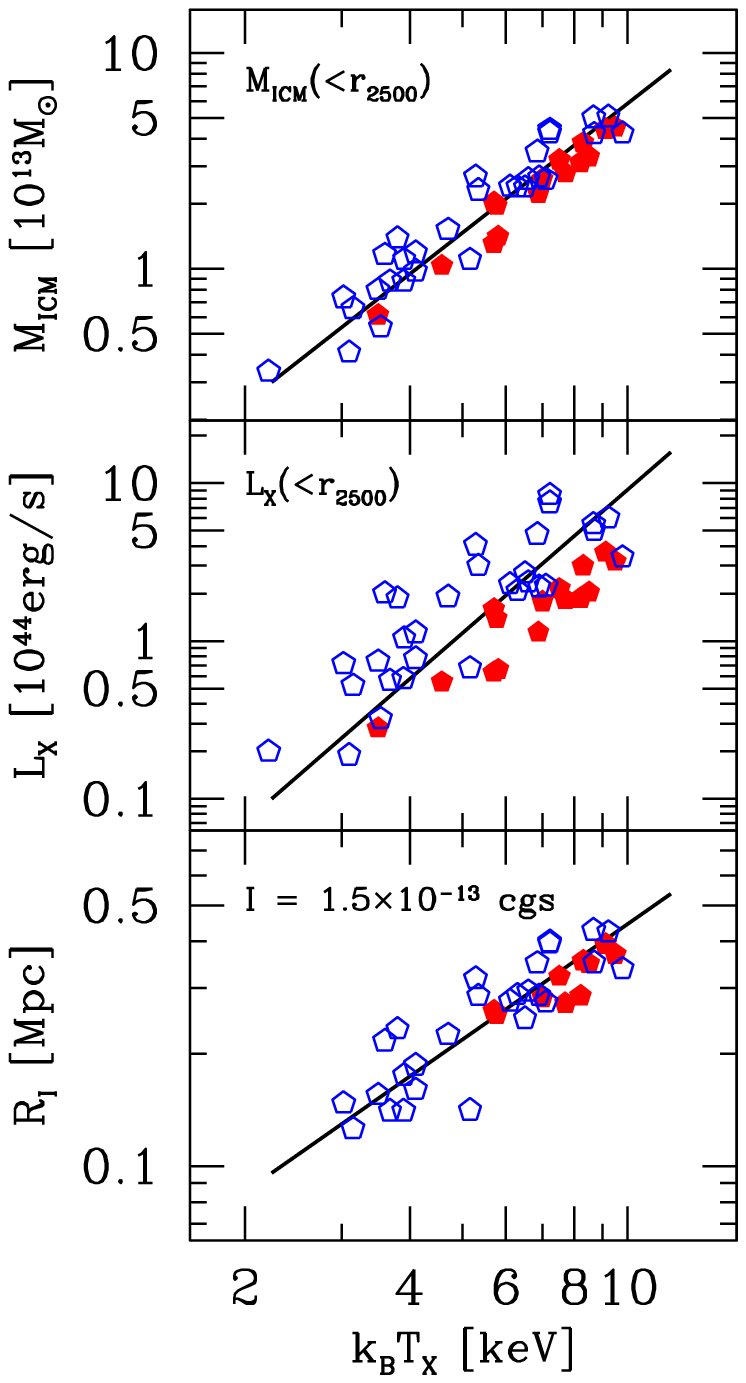}\end{minipage}
   \end{center}}
   \figcaption{\label{fig:scaling25} 
   Scaling relations for \Micm\ within \rtfive, X-ray luminosity projected within \rtfive, and isophotal size for an isophote of  $1.5\times10^{-13}$ erg \ps\ cm$^{-2}$ arcmin$^{-2}$, versus \Tx. The open and filled markers correspond to CC and NCC clusters, respectively.
        }
\end{inlinefigure}

All X-ray relations indicate a separation between CC and NCC clusters. The non-core-subtracted luminosities show the largest effect; this is not surprising, because the luminosity is the observable most affected by the buildup of dense gas in cool cluster cores. The relations that are more sensitive to the ICM distribution on larger scales show less of an offset between CC and NCC clusters.  In addition, these larger scale measurements (e.g., \Micmfive) show less scatter than relations at smaller scales (e.g., \Micmtfive). This is an interesting finding; it shows simply that core structural variations and temperature biases have a larger effect on relations measured in smaller regions around the core. This is an indication of the importance of core structure, which we discuss further below.

\subsection{$K$-band Light--Temperature Relation}

The X-ray observables provide information about the ICM.  We now turn to a very different cluster property, the $K$-band galaxy light. The near-infrared (NIR) light traces stellar mass better than optical bands, and is well correlated with such properties as cluster binding mass \citep[e.g.,][]{lin03b,lin04}. By comparing information from the X-ray and NIR observables, we hope to gain a better understanding of, for example, the true temperature bias introduced by cool cores. The $K$-band data we use is taken from \citet{lin04}; these data are available for 34 of the clusters in our sample. The measurements of near-IR light are effectively a simple sum of the light from individual galaxies, and thus represent a very different measure of galaxy cluster properties than the X-ray observables discussed above.

At the bottom of Figure~\ref{fig:scaling5} we plot the $K$-band luminosity projected within \rfive.   Here visual inspection suggests less evidence for a separation between the CC and NCC populations than in the X-ray relations. As with the X-ray observables, the scatter in the relation is dominated by the intrinsic scatter, as shown by the values for \sint\ and \sraw\ given in Table~\ref{tab:rawint}.

\section{Cool Core and Non-Cool Core Populations}
\label{sec:cores}

It is clear that the presence of cool cores affects all of the ICM-related scaling relations. As previously discussed, this is likely a result of various physical differences between CC and NCC clusters, which include a simple temperature bias effect.  Because we want to examine merger-related scatter about the scaling relations, we must account for this separation of populations in a way that will allow us to compare CC and NCC clusters in a manner independent of cooling effects.  One approach would be to treat these two populations individually, examining deviations from scaling relations in each.  In this section we effectively take this approach by applying a correction to the CC population so that in the mean these clusters lie on the same scaling relation as the NCC population.  Below we describe this approach as well as the amplitude of the offsets between the CC and NCC populations in each of the scaling relations. An alternative, more sophisticated approach will be presented in \S~\ref{sec:bright}.

\subsection{Aligning CC and NCC Populations}
\label{sec:align}

To begin with, we align the CC and NCC population scaling relations by scaling the mean temperature of all CC clusters by the same amount within a given relation. To find the appropriate scale factor for a relation, we increase \Tx\ for the CC clusters by a range of factors (using the same factor for all CC clusters), re-measure the relevant observable at each temperature, and measure the reduced $\chi^2$ for the entire sample at each scale factor. Because the measurement uncertainties are so small, we use a combination of measurement and intrinsic scatter, which reduces the dependence of the scale factor on outlier clusters. We measure $\chi^2$ versus temperature scale factor, then find a value for the intrinsic scatter \sint\ that makes the reduced $\chi^2=1$ at the $\chi^2$ minimum. We then find the temperature scale factor that corresponds to the new $\chi^2$ minimum, and so on, until the process converges. That is, for an observable $\mathcal{O}$, we find temperature scale factor $\lambda$ and intrinsic scatter \sint\ such that

\begin{equation}
\frac{\chi^2}{N_{\rm dof}} = \sum_i \frac{\left[\mathcal{O}_i(\lambda T_i)-\mathcal{O}_{\rm fit}(\lambda T_i)\right]^2}{\sigma_i^2 + \sigma_{\rm int}^2}=1
\end{equation}

\noindent
at the $\chi^2$ minimum for $N_{\rm dof}$ degrees of freedom.

 \begin{inlinefigure}
   {
   \begin{center}
   \epsfxsize=8.cm
   \end{center}}
   {\myputfigure{f4}{0}{.9}{-10}{-0}}
   \figcaption{\label{fig:LXerr_boost} 
   X-ray luminosity projected within \rfive\ ({\it top}), and deviation of clusters from the best fit relation ({\it bottom}), after scaling the CC cluster temperatures to align the CC and NCC cluster populations. Open and filled markers correspond to CC and NCC clusters, respectively. The uncertainties include both the measurement uncertainty in the luminosity and an effective temperature contribution to the luminosity uncertainty as described in the text.
     }
\end{inlinefigure}

Because aligning the CC and NCC cluster populations decreases the total scatter in each scaling relation, we must verify that the intrinsic scatter in the relations remains significantly greater than the measurement uncertainty. We again examine \LXfive, after scaling the CC cluster temperatures by a factor of 1.38, determined by the method described. The resulting scaling relation is shown in Figure~\ref{fig:LXerr_boost} (cf. the relation with non-scaled temperatures in Figure~\ref{fig:LXerr}). Although the total scatter in the relation is now significantly less than in the original relation, it still is clearly larger than the luminosity measurement uncertainties. We quantify this as before, measuring the raw and intrinsic scatter, and find that \sint$= 0.33$ and \sraw$=0.34$,  again showing the dominance of the intrinsic scatter over measurement uncertainties (scatter measurements for CC temperature-scaled relations are given in Table~\ref{tab:rawint}).

\begin{deluxetable}{cc}
\tablewidth{0pt}
\tabletypesize{\scriptsize}
\tablecaption{Cool Core Temperature Scale Factors}
\tablehead{
\colhead{Scaling relation} & \colhead{\Tx\ scale factor}  
}
\startdata
\LXtfive--\Tx &  $1.47 \pm 0.07$  \\
\LXfive--\Tx &  $1.38 \pm 0.06$  \\
\Micmtfive--\Tx &  $1.19 \pm 0.05$  \\
\LXCSfive--\Tx &  $1.15 \pm 0.04$  \\
\Rone--\Tx &  $1.10 \pm 0.07$  \\
\Micmfive--\Tx & $1.05\pm0.04$   \\
\Rthree--\Tx &  $1.01 \pm 0.05 $  \\
\LNIRfive--\Tx &  $ 0.77 \pm 0.08$  
\enddata

\label{tab:scale}
\end{deluxetable}

A plot of $\chi^2$ versus temperature scale factor for all of the X-ray and NIR scaling relations is shown in Figure~\ref{fig:chi}; the derived scale factors, plus their uncertainties, are listed in order from highest to lowest scale factor in Table~\ref{tab:scale}. Clearly, a greater scale factor is required to align the CC and NCC populations for parameters that measure a smaller, more core-dominated region of the cluster, as discussed in \S~\ref{sec:measure}. Similar scale factors of $\sim5\%$ and $\sim1\%$, respectively, will align the two populations for both \Micm\ within \rfive\ and \Rthree.
 There is a somewhat larger difference between temperature scale factors for \Micm\ within \rtfive, \Rone, and the core-subtracted luminosity within \rfive\ ($\sim19\%$, $\sim10\%$, and $\sim15\%$). A much greater scale factor is needed in either case for \LX\ ($\sim38\%$ within \rfive, $\sim47\%$ within \rtfive). This demonstrates that the luminosity is affected by more than just the shift in temperature due to the presence of emission from a cool core.

\begin{inlinefigure}
   {
   \begin{center}
   \epsfxsize=6.5cm
   \begin{minipage}{\epsfxsize}\epsffile{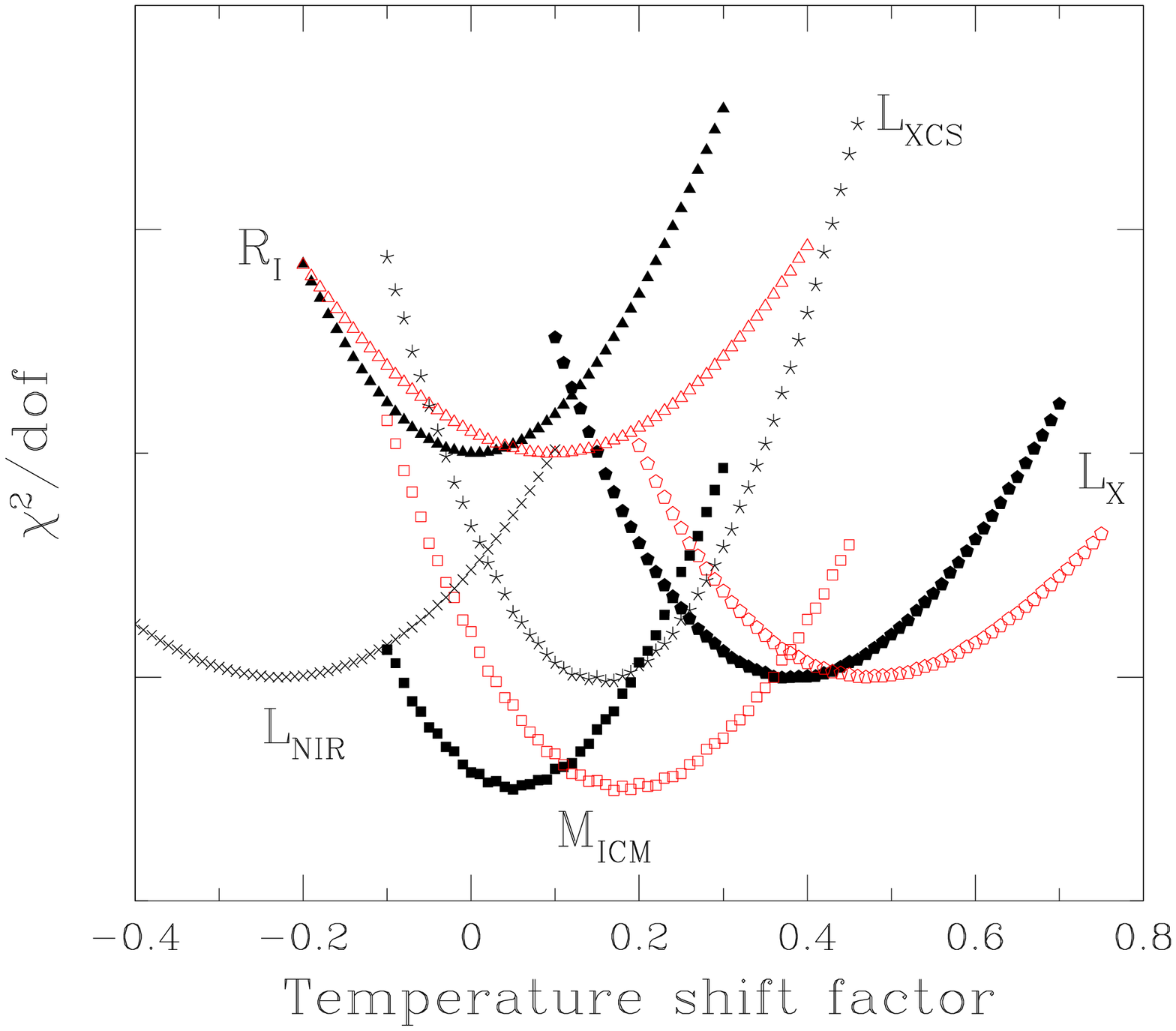}\end{minipage}
   \end{center}}
   \figcaption{\label{fig:chi} 
   Reduced $\chi^2$ versus CC temperature scale factor for each of the eight observable--temperature scaling relations. Pentagons: Projected luminosity within \rfive\ (filled) and \rtfive\ (open). Squares: ICM mass within \rfive\ (filled) and \rtfive\ (open). Triangles: Isophotal size for isophotes of $3\times10^{-14}$ erg \ps\ cm$^{-2}$ arcmin$^{-2}$ and $1.5\times10^{-13}$ erg \ps\ cm$^{-2}$ arcmin$^{-2}$ (open). Stars: Projected core-subtracted luminosity within \rfive. Crosses: Projected NIR light within \rfive. Vertical axis units are reduced $\chi^2$ value, but relations are offset vertically from one another for ease of viewing so vertical axis labels are not shown. 
     }
\end{inlinefigure}

\vskip20pt

\subsection{Examining the CC Temperature Bias}
\label{sec:RM}

Besides the cool core temperature bias, these scaling relations may be affected by cool core-related structural phenomena and merger-related phenomena that are in some way related to the presence or absence of cool cores.
A cool, dense core results in a sharp central brightness peak that drives up the total luminosity of the cluster, adding to the separation between CC and NCC clusters on the \LX--\Tx\ scaling relation. The increased central surface brightness also makes isophotal sizes larger and leads to higher measured gas masses in the central regions.  However, idealized studies of mergers of spherical, isolated clusters suggest that both ICM temperature and X-ray luminosity may be boosted during mergers, and that clusters can be perturbed from scaling relation by the merger \citep{ricker01}.  If present, this behavior would also tend to separate the merging and non-merging (and hence NCC and CC) populations.  

\begin{inlinefigure}
   {
   \begin{center}
   \epsfxsize=8.cm
   \begin{minipage}{\epsfxsize}\epsffile{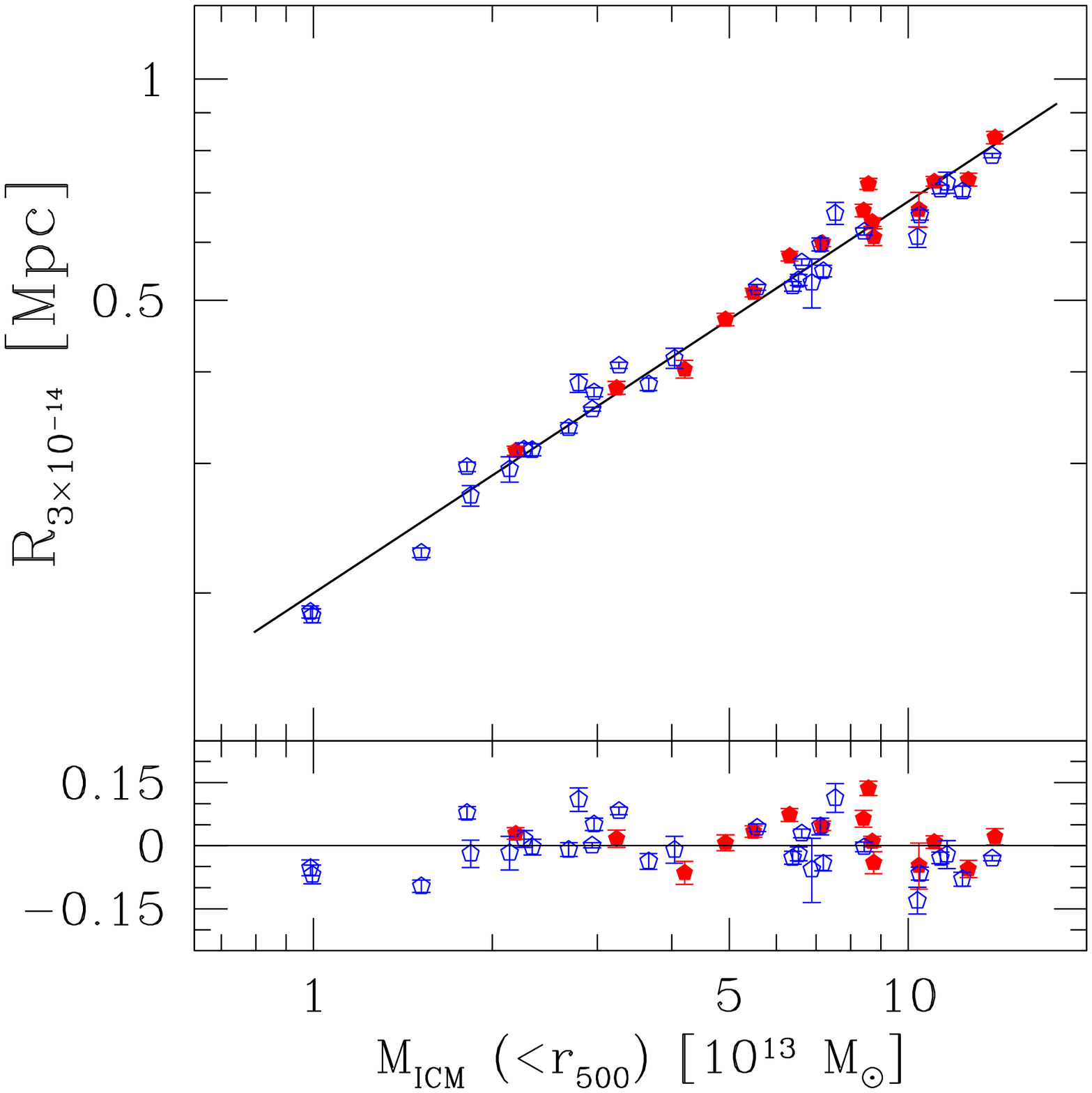}\end{minipage}
   \end{center}}
   \figcaption{\label{fig:RM500} 
   \Rthree\ plotted versus \Micmfive, along with the best-fit relation ({\it top}), and deviation in \RI\ from the best-fit relation ({\it bottom}). Open and filled markers correspond to CC and NCC clusters, respectively. Uncertainties are a combination of \RI\ measurement uncertainties and an effective uncertainty in \RI\ due to the mass uncertainty obtained using the scaling relation slope, as was done with temperature uncertainties for other relations.
     }
\end{inlinefigure}

A good way of differentiating between these two effects is to examine a relation that has minimal  temperature-dependence.  To examine the impact of cool core effects on scaling relations at large radii, we plot \Rthree\ versus \Micmfive\ as shown in Figure~\ref{fig:RM500}.  Isophotal size is independent of temperature, and at a low isophote core effects should be of little importance. The ICM mass within a large radius will have only a slight dependence on core features, because only a small fraction of the cluster ICM mass lies within the core region.  Though \rfive\ depends on temperature, any bias effect this introduces is weak.  Hence, a scaling relation composed of these two observables provides a test of how much the cool core bias affects the structure outside the cluster core. Indeed, the relation shows no particular suggestion of separation between the CC and NCC populations, and the total scatter in the relation is much smaller than for any of the previously discussed observable--temperature relations. This provides evidence that the primary contributor to the CC/NCC separation in the observables at large radii is indeed a simple temperature bias, and not structural changes related to the development or disruption of cool cores.  This also gives us confidence that the offset between the CC and NCC populations is caused by the onset of a cooling instability within the cluster core rather than by shock-induced temperature and structural changes during mergers, which we would expect to be most apparent in observables that are sensitive to cluster structure outside the core (where relaxation timescales are the longest).

Adopting this perspective, we can take the scale factors for relations involving observables that are less core sensitive to estimate the scale of the temperature bias.  For example, the \Micmfive\ and \Rthree\ relations have scale factors of 1.04$\pm$0.04 and 1.01$\pm$0.05, respectively, suggesting that temperature biases are at the few percent level, and that it is indeed structural differences in the core that are driving the larger scale factors seen in the more core sensitive observables.  Interestingly, the near-IR relation shows less evidence for a cool core-related separation. In fact, following the same procedure as for the X-ray relations indicates that a {\it negative} scale factor of $\sim$23\% is required to align the CC and NCC poplulations. (This may be partly driven by a few outliers, but having no reason to discard these data points, we do not exclude them.) This suggests that the galaxy population in CC clusters contains systematically less light than the galaxy population in NCC clusters, a result that deserves further attention.

 \begin{deluxetable*}{cccccccc}
\tablewidth{0pt}
\tablecaption{Intrinsic Scatter CC and NCC Subsamples}
\tablehead{
& \multicolumn{3}{c}{With CC Temperature Scaling}  & &  \multicolumn{3}{c}{3 Paramter (\OTI)} \cr
\cline{2-4} \cline{6-8}
\colhead{Scaling Relation} & \colhead{CC \sint} & \colhead{NCC \sint} & \colhead{Diff. (\%)\tablenotemark{a}} & & \colhead{CC \sint} & \colhead{NCC \sint} & \colhead{Diff. (\%)\tablenotemark{a}}
}
\startdata
\Micmfive--\Tx & 0.20 & 0.09 & 99.6\,+ && 0.17 & 0.11 & 93.0\,+  \\
\Micmtfive--\Tx & 0.22 & 0.10 & 99.6\,+ && 0.17 & 0.08 & 99.7\,+ \\
\LXfive--\Tx & 0.37 & 0.28 & 73.0\,+ && 0.28 & 0.18 & 90.7\,+ \\
\LXtfive--\Tx &  0.44 & 0.31 & 83.6\,+ && 0.40 & 0.20 & 99.3\,+ \\
\LXCSfive--\Tx &  0.24 & 0.24 & 5.3\,+ && 0.22 & 0.22 & 3.6\,- \\
\Rthree--\Tx &  0.15 & 0.10 & 91.9\,+ && 0.14 & 0.11 & 62.3\,+\\
\Rone--\Tx &  0.19 & 0.10 & 95.1\,+ && 0.16 & 0.10 & 84.0\,+ \\
\LNIRfive--\Tx & 0.21 & 0.09 & 99.3\,+ && 0.23 & 0.12 & 97.3\,+ 

\enddata
\tablenotetext{a}{Percent likelihood that scatter measurements for CC and NCC subsamples are different; see text. Plus sign indicates that CC scatter value is higher than NCC scatter value.}
\label{tab:ccsplit}
\end{deluxetable*}

\subsection{Intrinsic Scatter in CC and NCC Populations}

Having removed to first order, via temperature scaling of CC clusters, the separation between cluster populations, we can begin to study the effects of merging on scatter about scaling relations; because mergers are expected to disrupt cool cores, and because merging clusters are naively expected to have greater scatter about scaling relations, one would expect to observe greater scatter in NCC clusters. We therefore measure the intrinsic scatter in the temperature-scaled CC population and the NCC population separately as a test of overall structural differences between them. That is, we measure the scatter of each (CC and NCC) population about the {\it same} best-fit scaling relation. These values are given for all eight \OT\ scaling relations in Table~\ref{tab:ccsplit}. We use an F-test to quantify the significance of differences between CC and NCC scatter for a given relation; the table lists the percent significance level at which equality of the variances is rejected.  Remarkably, we do not observe greater scatter in the NCC population; indeed, the CC population has significantly (i.e., $>68\%$ significance) greater scatter than the NCC population in all but one scaling relation.

The uniform CC temperature scaling method treats all CC clusters in exactly the same way. The greater scatter in the CC population, however, provides evidence of significant structural variation within that population---more variation, in fact, than in the NCC population. We are thus motivated to find a method to reduce CC/NCC separation that takes into account the variability of individual cluster structure. In the next section we present such a method.

\section{Peak Surface Brightness as a Measure of Cool Core Strength}
\label{sec:bright}

 In this section we discuss another method of reducing cooling-related scatter in scaling relations: the use of peak surface brightness as an indication of cool core ``strength".  We include surface brightness as a third parameter in scaling relations, show that this parameter's contribution to the relation is significant in all X-ray observable--temperature relations, and demonstrate its usefulness in reducing cool core-related scatter.

\begin{inlinefigure}
   {
   \begin{center}
   \epsfxsize=8.cm
   \begin{minipage}{\epsfxsize}\epsffile{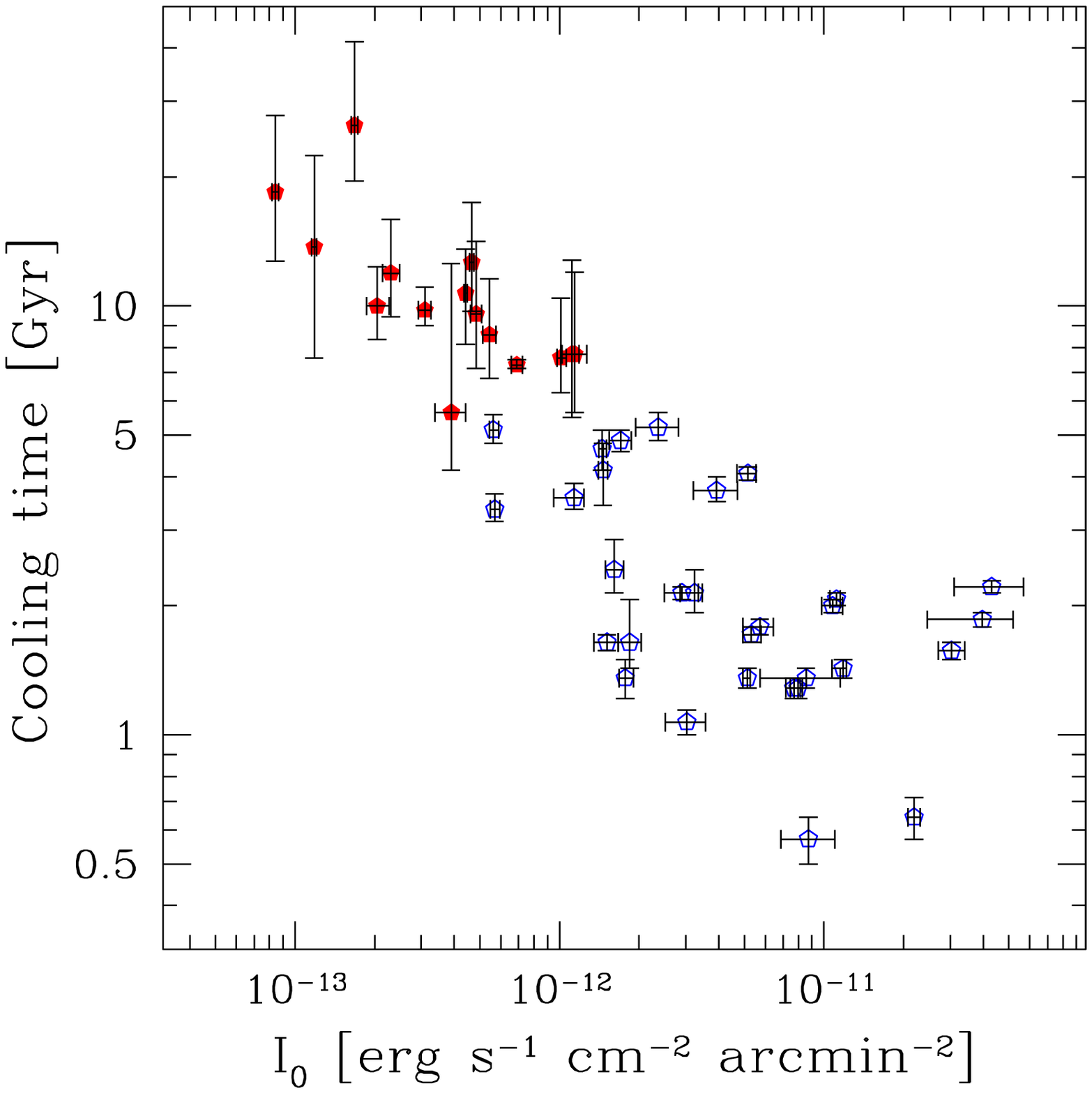}\end{minipage}
   \end{center}}
   \figcaption{\label{fig:coolI0} 
  Central cooling time from \citet{peres98} plotted versus peak surface brightness \Io. Open and filled markers correspond to CC and NCC clusters, respectively.
     }
\end{inlinefigure}

\subsection{Brightness Measurements}

We use measurements of the peak surface brightness \Io\ from MME. These values were obtained by fitting $\beta$ models to azimuthally-averaged cluster surface brightness profiles. Clusters that appeared relaxed and displayed residuals consistent with a central emission excess  were fit with a double $\beta$ model, i.e., two models with the same $\beta$ but different \Io\ and core radius were summed and fitted together; the cluster \Io\ is then the sum of the individual \Io\ from each model.  In Figure~\ref{fig:coolI0} we plot the central cooling time versus \Io\ for the 45 clusters in our sample. There is a clear correlation between the two quantities, which is not surprising as the central cooling time is derived from the central surface brightness profile.  Note that the PSF corrected central surface brightness varies by a factor of $\sim$500 for our flux limited cluster sample, suggesting that even in the low signal to noise regime it should be possible to differentiate the low and high surface brightness systems.

\begin{inlinefigure}
   {
   \begin{center}
   \epsfxsize=8.cm
   \begin{minipage}{\epsfxsize}\epsffile{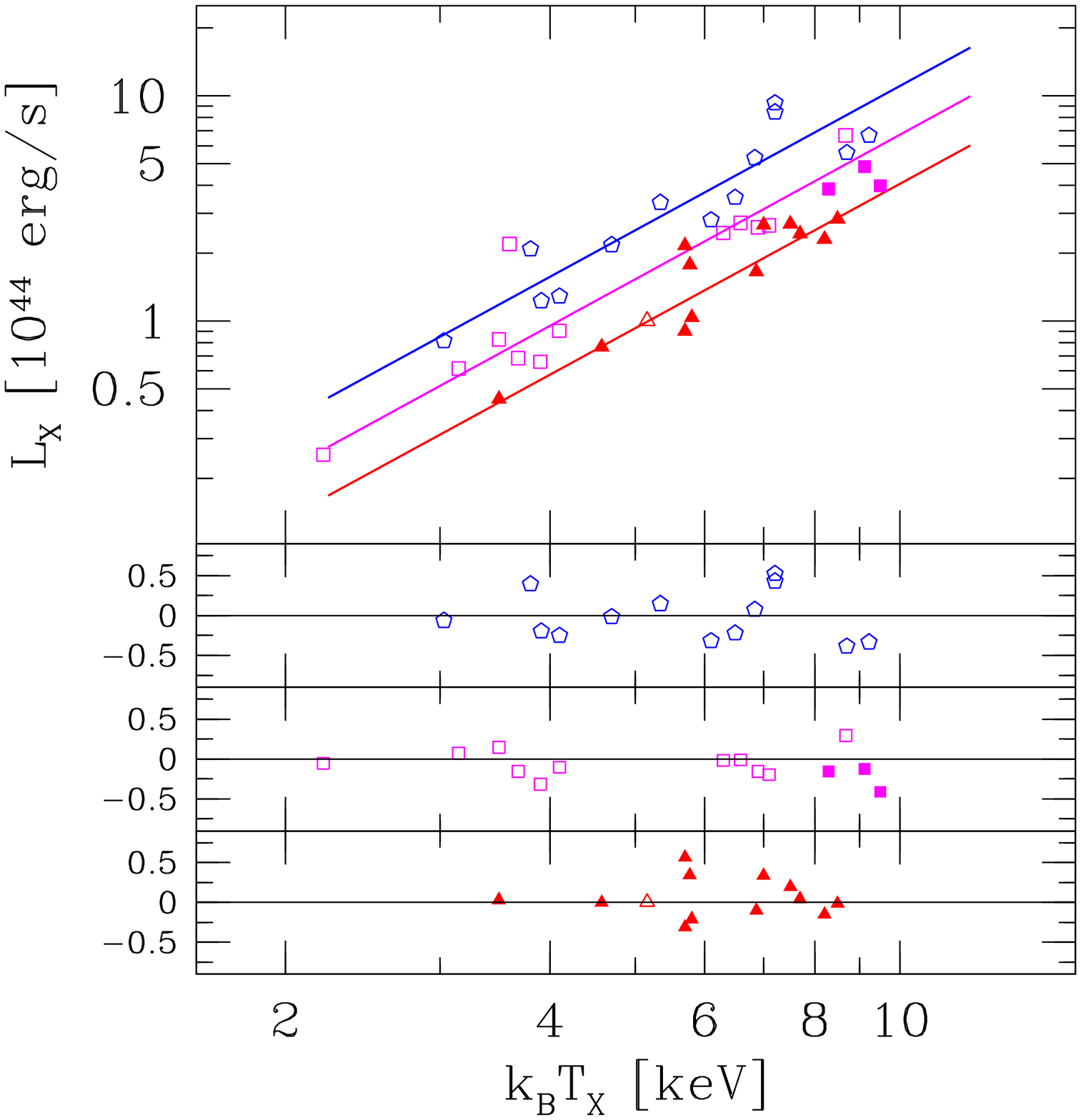}\end{minipage}
   \end{center}}
   \figcaption{\label{fig:LTI} 
  X-ray luminosity projected within \rfive\ plotted versus temperature ({\it top}), 
  and deviation of clusters from the best fit \LX--\Tx--\Io\ relation calculated for three values of \Io\ ({\it bottom}). For this plot the cluster sample was divided into three subsamples based on \Io; the pentagons, squares, and triangles are the clusters in the subsamples with the highest, middle, and lowest values of \Io, respectively. Open and filled markers correspond to CC and NCC clusters, respectively. 
     }
\end{inlinefigure}

\begin{deluxetable*}{cccc}
\tablewidth{0pt}
\tablecaption{Best Fit Temperature and Brightness Scaling Parameters}

\tablehead{
\colhead{Observable} & \colhead{\Tx\ dependence}  & \colhead{\Io\ dependence} & \colhead{\Io\ dependence}  \\
 & & & \colhead{(\Tx\ bias removed)\tablenotemark{a}}
 }
\startdata
\Micmfive & $1.94 \pm 0.09$ & $0.06 \pm 0.02$ & $0.00 \pm 0.03$ \\
\Micmtfive & $1.92 \pm 0.08$ & $0.10 \pm 0.02$ & $0.04 \pm 0.03$\\
\LXfive  & $2.13 \pm 0.10$ & $0.26 \pm 0.03$ & $0.20 \pm 0.04$ \\
\LXtfive & $2.33 \pm 0.18$ & $0.31 \pm 0.03$ & $0.24 \pm 0.04 $ \\
\LXCSfive &  $2.21 \pm 0.10$ & $0.10 \pm 0.03$ & $0.03 \pm 0.04 $ \\
\Rthree & $1.03 \pm 0.07$ & $0.03 \pm 0.01$ & $0.00 \pm 0.01$ \\
\Rone &  $1.02 \pm 0.08$ & $0.07 \pm 0.02$ & $0.04 \pm 0.02$ \\
\LNIRfive &  $1.41 \pm 0.14$ & $0.00 \pm 0.03$ & $-0.04 \pm 0.03 $
\enddata
\tablenotetext{a}{That is, the factor $\gamma$ as defined in Eq.~\ref{eq:OIT}.}
\label{tab:scaling}

\end{deluxetable*}

\subsection{Observable--Temperature--Brightness Relations}

We now test whether \Io\ is a significant parameter by including it in the observable--temperature scaling relations. That is, for each observable $\mathcal{O}$ we assume a functional form $\mathcal{O} \propto T_{\rm X}^\alpha I_0^\beta$ and solve for $\alpha$, $\beta$, and the normalization.  We first examine the X-ray luminosity projected within \rfive. A plot of \LXfive\ versus \Tx\ is shown in Figure~\ref{fig:LTI}. In this figure we have divided the cluster sample into three subsamples based on \Io, and we plot the measured \LX--\Tx--\Io\ relation for a value of \Io\ in the middle of each subsample; we  also show the deviation in luminosity for each subsample.  This plot shows qualitatively that the scatter about the \LX--\Tx--\Io\ is much smaller than about the \LX--\Tx\ relation (c.f. the \LX--\Tx\ relation in Fig.~\ref{fig:LXerr}).
Quantitatively, we find that the \LXfive--\Tx--\Io\  relation has a best-fit power-law dependence on \Io\ with an exponent of $0.26 \pm 0.03$ (uncertainty is obtained from bootstrap resampling and refitting); the \Io\ dependence is thus indeed significant. The raw and intrinsic scatter in \LXfive\ about the relation are 0.26 and 0.24, respectively, much lower than the corresponding values for the original \LX-\Tx\ relation  and also lower than for the \LX-\Tx\ relation with CC temperature shifting (see Table~\ref{tab:rawint}).  This dramatic reduction in scatter in the luminosity-related scaling relations suggests that it should be possible to use luminosity together with central surface brightness as a much more accurate cluster mass estimator than luminosity alone.

\OTI\ scaling relation exponents for the rest of the relations are given in Table~\ref{tab:scaling}. All X-ray observables  have a dependence on \Io\ that is significant at greater than two standard deviations; 
\LNIR\ does not show any \Io\ dependence. We conclude that the peak surface brightness does provide us with useful information about cluster structure. Its usefulness for our present purpose is clear given the extent to which scatter about scaling relations is reduced by its introduction. Table~\ref{tab:rawint} gives measured raw and intrinsic scatter values for each of the eight observable--\Tx--\Io\ relations, and  separate measurements of CC and NCC scatter are given in Table~\ref{tab:ccsplit}. For X-ray observables we find reduced scatter in the $\mathcal{O}$--\Tx--\Io\ relations compared to the original $\mathcal{O}$--\Tx\ relations, and generally less than for the temperature-scaled relations. The exception is \LNIR\ which, having no dependence on \Io, does not show reduced scatter compared to the original \LNIR--\Tx\ relation.

There have been several recent studies of the evolution of cluster scaling relations at intermediate and high redshift. Some studies have found positive evolution of cluster X-ray luminosity, consistent with that expected from self-similarity arguments \citep[e.g.,][]{vikhlinin02,maughan05}, but other observations are consistent with zero or somewhat negative evolution of \LX--\Tx\ and \Micm--\Tx\ scaling relation normalizations \citep[e.g.,][]{borgani01,holden02,ettori04a}. Given the difference in normalizations for the CC and NCC populations, this may be partially explained by a simple change in the cool core fraction in the samples being compared, which will shift the normalization of the entire cluster population. It would thus potentially be useful to compare $\mathcal{O}$--\Tx--\Io\ relations at low and high redshift, as the inclusion of a parameter measuring cool core strength may reduce this evolution effect.

\subsection{ Temperature and Other Observable Biases}
\label{sec:bias}

We can use this three parameter scaling relation approach to again estimate the scale of the temperature biases.  Scaling relations involving properties that have low dependence on core structure should have little to no dependence on \Io. That all X-ray observable--temperature relations show a significant \Io\ dependence can be taken as evidence for temperature biases in CC clusters; that is, the cool gas in CC cluster cores biases emission-weighted mean temperatures so that CC clusters appear to lie above scaling relations. We can attempt to quantify this bias in terms of \Io. We assume scaling relations of the form

 \begin{equation}
 \label{eq:OIT}
 \mathcal{O} = \mathcal{O}_0 T_{\rm X}^\alpha I_0^\beta = \mathcal{O}_0 \left( \lambda(I_0) T_{\rm X} \right)^\alpha I_0^\gamma
 \end{equation}
 
\noindent for each observable $\mathcal{O}$. If a scaling relation has no intrinsic dependence on \Io\ (i.e., $\gamma=0$), then $\lambda(I_0) \propto I_0 ^{\beta / \alpha}$. We assume that the \Micmfive--\Tx\ and \Rthree--\Tx\ relations have no intrinsic \Io\ dependence (see \S~\ref{sec:RM}), and so $\lambda(I_0) \propto I_0 ^{(0.03 \pm 0.01)}$. The temperature bias thus varies by $\sim$20\% over the range of \Io\ in our sample; using the median values of \Io\ for the CC and NCC populations ($4.5 \times 10^{-14}$ and $4.4\times10^{-13}$ erg s$^{-1}$ cm$^{-2}$ arcmin$^{-2}$, respectively)
gives an average CC temperature scale factor  of $1.07 \pm 0.02$ (simply averaging the CC temperatures scale factors for \Micmfive\ and \Rthree\ from \S~\ref{sec:align} gives $1.03 \pm 0.03$).

For the other scaling relations, $\gamma = \beta - (0.03 \pm 0.01)\alpha $. The values for $\gamma$, i.e., the dependence on \Io\ when the temperature bias is removed, are given in the fourth column of Table~\ref{tab:scaling}. Not surprisingly, the X-ray luminosities still have by far the strongest \Io\ dependence, which differs by a factor of four between the highest and lowest values of \Io\ in our sample. The \Micmtfive\ dependence on \Io\ varies by $\sim$30\% over the sample.  This suggests that the gas fraction varies significantly within \rtfive\ depending on the strength of the cool core;   variations of this scale will present challenge to those using cluster gas fractions of ``relaxed" clusters to precisely study cosmology.

\vspace{3pt}

\section{Substructure as a Source of Scatter}
\label{sec:subscatter}

We now examine the relationship between substructure and the position of clusters on scaling relations. Having introduced two ways to remove the cool core temperature bias, we now attempt to examine the merger related structural differences as deviations by clusters from scaling relations. In this section, we review the substructure measurements, and then discuss how deviation from scaling relations depends on substructure. We then compare results from the observational sample to an ensemble of hydrodynamical simulations.

\subsection{Substructure Measurements}  
\label{sec:substructure}

High-resolution instruments such as \Chandra\ reveal hydrodynamic phenomena such as cold fronts that are clearly related to merging. However, it is generally not necessary to directly observe such features to find evidence of merger-related activity. Relatively crude, low-order moments of the X-ray surface brightness distribution such as centroid variation and ellipticity have been shown to be effective at separating clusters with recent major mergers from more relaxed systems \citep[][]{mohr93,evrard93}, although these measurements are essentially unaffected by mergers along the line of sight.  These measurements do not require the high resolution of \Chandra, and they were used to show that more than half of clusters display substructure in their ICM \citep{mohr95}.

The centroid variation $w$ is a measure of the ``center shift", or skewness of the photon distribution of a cluster.  There are many ways to measure $w$; here, we measure within an isophote of $2\times10^{-14}$ erg \ps\ cm$^{-2}$ arcmin$^{-2}$, chosen as the lowest isophote that can be used for all of our cluster images.  We measure the centroid of the portion of the cluster that has surface brightness above this isophote.  We then examine the cluster at steadily brighter isophotes and measure the variance in the centroids measured for all these isophotes.  The centroid variation $w$ is the square root of this variance. We scale $w$ to be in units of \rfive\ for each cluster; using the fractional variation rather than the raw variation simply accounts for the fact that the size of cluster virial radii can vary by as much as a factor of $\sim$4 from low mass to high mass systems.  
Measuring the centroid variation using regions defined by isophotes rather than using circular annuli as in \citet{mohr93} can provide information that circular apertures may not;  during mergers, clusters often do not have circular surface brightness distributions, and elongated structure may be missed or underemphasized by fixed circular apertures.

We determine the axial ratio $\eta$ from the flux-weighted second moments of the photon distribution, using an aperture centered on the brightness peak. That is, we measure moments
\begin{equation}
\label{eq:matrix}
M_{i j} = \sum I  x_i x_j ~,
\end{equation}
\noindent where the sum is carried out over all pixels within a chosen aperture, $x_i$ are pixel coordinates  ($x$ or $y$) relative to the center of the aperture, and $I$ is the measured intensity in the pixel. We measure $\eta$ within an aperture of radius \rfive\ for each cluster, except for four clusters for which \rfive\ either is larger than our PSPC image or is close enough to the edge that background problems arise.  In these cases we use apertures of radius \rtfive. Using virial radii for the apertures provides a more physically meaningful scale for examining substructure than using a fixed metric radius.  Diagonalizing the matrix obtained from equation~(\ref{eq:matrix}) gives the lengths of the major and minor axes, from which we then obtain the axial ratio.  While high ellipticity is not a certain indicator of cluster substructure, and is sometimes observed even in apparently relaxed clusters \citep[e.g.,][]{schuecker01},  hydrodynamical simulations show	 that during major mergers the ICM is typically highly flattened \citep{evrard93,pearce94}.

\begin{inlinefigure}
   {
   \begin{center}
   \epsfxsize=8cm
   \begin{minipage}{\epsfxsize}\epsffile{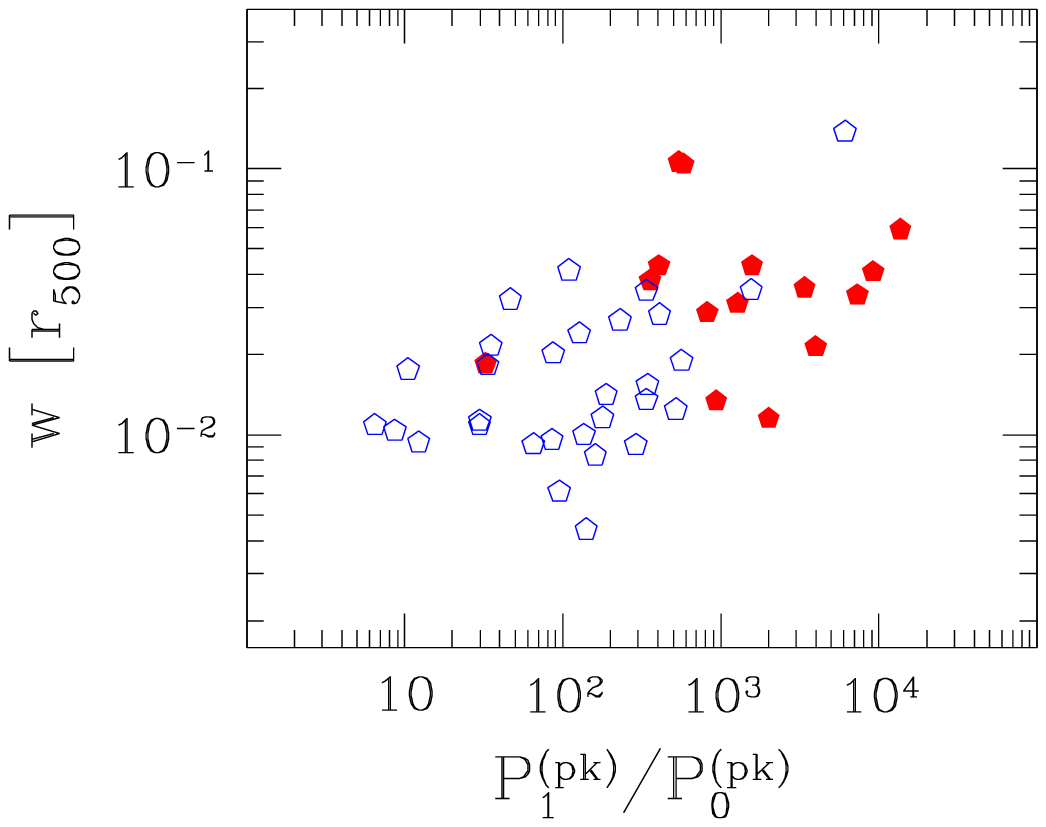}\end{minipage}
   \end{center}}
   \figcaption{\label{fig:substructure1} 
   Comparison of centroid variation $w$ to the power ratio \Pone\ for the 45 clusters in the PSPC sample. Power ratio values are quoted in units of $10^{-7}$; centroid variation is plotted in units of \rfive. Open and filled markers are CC and NCC clusters, respectively.
        }
\end{inlinefigure}

Another method to quantify cluster substructure makes use of ``power ratios"  \citep{buote95}. This involves measuring moments of the surface brightness distribution $\Sigma$ within some radius $R_{\rm ap}$:
\begin{equation}
\label{eq:moments1}
a_m(R_{\rm ap})=\int_{R^\prime\leq R_{\rm ap}} \Sigma({\bf x}^\prime)(R^\prime)^m\cos m\phi^\prime \, d^2x^\prime ~,
\end{equation}
\begin{equation}
\label{eq:moments2}
b_m(R_{\rm ap})=\int_{R^\prime\leq R_{\rm ap}} \Sigma({\bf x}^\prime)(R^\prime)^m\sin m\phi^\prime \, d^2x^\prime ~.
\end{equation}
\noindent The ``powers" $P_m$ are then given by:
\begin{equation}
\label{eq:P0}
P_0=[a_0 \ln (R_{\rm ap})]^2 
\end{equation}
\noindent for $m=0$, and
\begin{equation}
\label{eq:Pm}
P_m=\frac{1}{2m^2 R_{\rm ap}^{\,2m}}(a_m^2+b_m^2) 
\end{equation}
\noindent for $m > 0$, where $R_{\rm ap}$ is the radius of the circular aperture in which the moments are measured. The quantities of interest are the ratios $P_m/P_0$; the division by $P_0$ normalizes the flux within the radius of interest, allowing comparison of cluster observations with different fluxes and exposure times. The quantity \Pone, which is calculated within an aperture centered on the cluster surface brightness peak, is similar to the centroid variation $w$. Other ratios $P_{m>1}/P_0$ are measured within an aperture centered on the point where the centroid variation is at a minimum; the quantity \Ptwo\ is related to the axial ratio $\eta$.

Using a sample of PSPC observations, \citet{buote96} argue that certain relationships between power ratios may be viewed as evolutionary tracks. We seek here to find correlations between power ratios---mainly the simplest to interpret, \Pone\ and \Ptwo---and the deviations of clusters from scaling relations. We break from their approach of using fixed metric radii and use an aperture that scales with the cluster mass or temperature; specifically, we study the power ratios within the same characteristic radii (\rfive\ or \rtfive) as we do with axial ratios. This provides a more physically meaningful scale for a cluster sample that spans more than an order of magnitude in mass.   Figure~\ref{fig:substructure1} contains a plot of centroid variation versus \Pone\ and Figure~\ref{fig:substructure2} contains a plot of axial ratio versus \Ptwo. The centroid variation $w$ and the axial ratio $\eta$ are correlated with the primary power ratio \Pone\ and \Ptwo, respectively.  Because  neither pair of substructure measurements is perfectly correlated, we benefit from using all four measurements.

\begin{inlinefigure}
   {
   \begin{center}
   \epsfxsize=8.cm
   \begin{minipage}{\epsfxsize}\epsffile{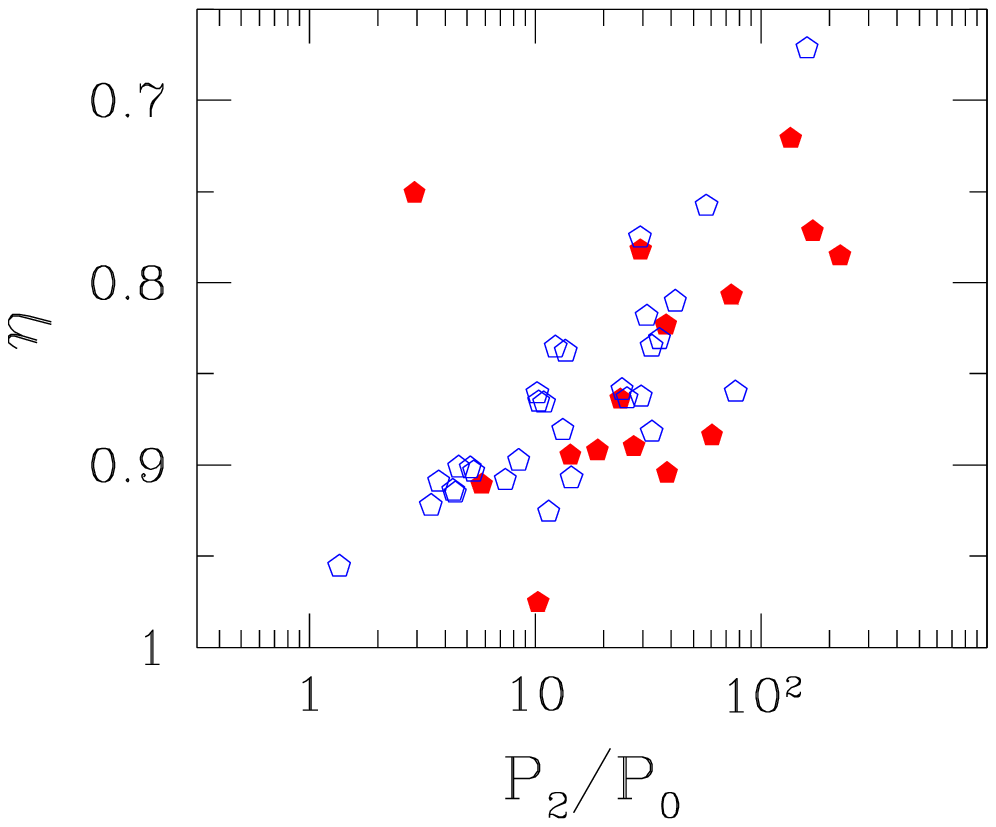}\end{minipage}
   \end{center}}
   \figcaption{\label{fig:substructure2} 
   Comparison of axial ratio $\eta$ to the power ratio \Ptwo\ for the 45 clusters in the PSPC sample. Power ratio values are quoted in units of $10^{-7}$. Note that axial ratio is plotted from highest to lowest, i.e., (vertically) higher points have greater ellipticity. Open and filled markers are CC and NCC clusters, respectively.
        }
\end{inlinefigure}

\begin{deluxetable*}{ccccccccccccc}
\tablewidth{0pt}
\tabletypesize{\scriptsize}
\tablecaption{Intrinsic Scatter in Scaling Relations, Split By Substructure}
\tablehead{
Scaling Relation & \multicolumn{12}{c}{\sint} \cr
\cline{2-13}
& \multicolumn{3}{c}{ Split by $w$} &
\multicolumn{3}{c}{ Split by $\eta$} &
\multicolumn{3}{c}{ Split by \Pone} &
\multicolumn{3}{c}{ Split by \Ptwo}  \\
& \colhead{L} &  \colhead{H} & \colhead{Diff.(\%)\tablenotemark{a}} & 
 \colhead{L} & \colhead{H} & \colhead{Diff.(\%)\tablenotemark{a}} &
 \colhead{L} &  \colhead{H} & \colhead{Diff.(\%)\tablenotemark{a}} &
 \colhead{L} &  \colhead{H} & \colhead{Diff.(\%)\tablenotemark{a}} 
 }
\startdata
\multicolumn{13}{c}{Observational sample} \\
\Micmfive--\Tx--\Io &  0.18 & 0.12 & 95.5\,+ & 0.18 & 0.13 & 79.2\,+ &  0.15 & 0.15 & 2.5\,- & 0.15 & 0.15 & 8.4\,+ \\
\Micmtfive--\Tx--\Io & 0.18 & 0.11 & 96.1\,+ & 0.15 &0.15 & 4.6\,- & 0.14 & 0.15 & 17.9\,- & 0.14 & 0.16 & 48.9\,- \\
\LXfive--\Tx--\Io &  0.32 & 0.18 & 98.4\,+ & 0.16 & 0.31 & 99.4\,- & 0.31 & 0.17 & 99.1\,+ & 0.30 & 0.19 & 94.9\,+ \\
\LXtfive--\Tx--\Io & 0.45 & 0.19 & 100.0\,+ & 0.32 & 0.38 & 57.8\,- & 0.38 & 0.31 & 61.7\,+ & 0.41 & 0.28 & 91.6\,+ \\ 
\LXCSfive--\Tx--\Io & 0.26 & 0.18 & 86.2\,+ &  0.19 & 0.24 & 69.9\,- & 0.25 & 0.19 & 77.4\,+ & 0.24 & 0.20 & 57.7\,+ \\
\Rthree--\Tx--\Io & 0.16 & 0.10 & 93.1\,+ &  0.14 & 0.12 & 57.6\,+ & 0.12 & 0.14 & 51.9\,- & 0.14 & 0.12 & 42.4\,+ \\
\Rone--\Tx--\Io &  0.15 & 0.14 & 21.1\,+ & 0.09 & 0.17 	& 98.3\,-  & 0.13 & 0.16 & 56.6\,- & 0.13 & 0.16 & 65.4\,- \\
\LNIRfive--\Tx--\Io & 0.20 & 0.21 & 16.8\,- & 0.18 & 0.22 & 57.0\,- & 0.21 & 0.20 & 18.1\,+ & 0.19 & 0.22 & 43.9\,+ \\
\Micmfive--\Rthree & 0.06 & 0.05 & 37.1\,+ & 0.05 & 0.06 & 28.8\,- & 0.05 & 0.06 & 82.3\,- & 0.05 & 0.06 & 8.7\,- \\
\LXCSfive--\Rthree & 0.05 & 0.05 & 10.5\,- & 0.05 & 0.06 & 28.2\,- & 0.05 & 0.05 & 37.6\,- & 0.05 & 0.05 & 15.1\,- \\
\LNIRfive--\Rthree & 0.19 & 0.14 & 72.2\,+ &	 0.20 & 0.14 & 82.0\,+ & 0.16 & 0.17 & 14.7\,- & 0.17 & 0.16 & 17.8\,+ \\
\multicolumn{13}{c}{Simulated cluster sample}\\
\Micmfive--\Tx 	& 0.21 & 0.21 & 6.0\,+   & 0.21 & 0.20 & 10.9\,+ & 0.24 & 0.17 & 87.4\,+ & 0.22 & 0.19 & 39.5\,+ \\
\LXCSfive--\Tx 	& 0.26 & 0.30 & 56.1\,- & 0.25 & 0.30 & 59.3\,- & 0.23 & 0.31 & 78.1\,- & 0.27 & 0.29 & 29.4\,-  \\
\RI--\Tx 		& 0.09 & 0.11 & 62.6\,- & 0.10 & 0.11 & 54.8\,- & 0.11 & 0.10 & 12.4\,+ & 0.10 & 0.11 & 26.4\,- 
\enddata
\tablecomments{ \ L and H are low substructure (low $w$, \Pone, and \Ptwo, and high $\eta$) and high substructure, respectively. }
\tablenotetext{a}{Percent likelihood that scatter measurements for low and high substructure subsamples are different; see text. Plus sign indicates that low substructure sample scatter is higher; minus sign indicates that high substructure sample scatter is higher.}
\label{tab:subsplit}

\end{deluxetable*}

\subsection{Substructure and Scaling Relations: CC Temperature Scaling versus \OTI\ Relations}
\label{sec:subsscaleOTI}

To test for merger-related scatter in scaling relations, we measure the scatter about scaling relations by subsamples of clusters, split according to the four substructure measures discussed above. 
We wish to minimize  cool core related scatter, and we have discussed two methods of doing so in this paper: uniform CC cluster temperature scaling, and use of peak surface brightness as a third parameter in observable--temperature scaling relations.

As shown above, the \OTI\ relations generally have lower scatter than the CC temperature-scaled relations; however, one may wonder whether scatter information is being lost in the \OTI\ relations. Qualitative comparisons suggests that this is not the case. We give one example here: in Figure~\ref{fig:diffLx} we plot the difference between the data and best-fit \LXfive--\Tx--\Io\ relation and the best-fit CC temperature-scaled \LXfive--\Tx\ relation versus two substructure measures, centroid variation and axial ratio. It is clear that the \LX--\Tx--\Io\ relation is not masking any increase in scatter in high-substructure clusters. Similar results are seen for other scaling relations; we thus choose to focus on the \OTI\ scaling relations for our study of merger-related scatter below.

\begin{inlinefigure}
   {
   \begin{center}
   \epsfxsize=9.cm
   \begin{minipage}{\epsfxsize}\epsffile{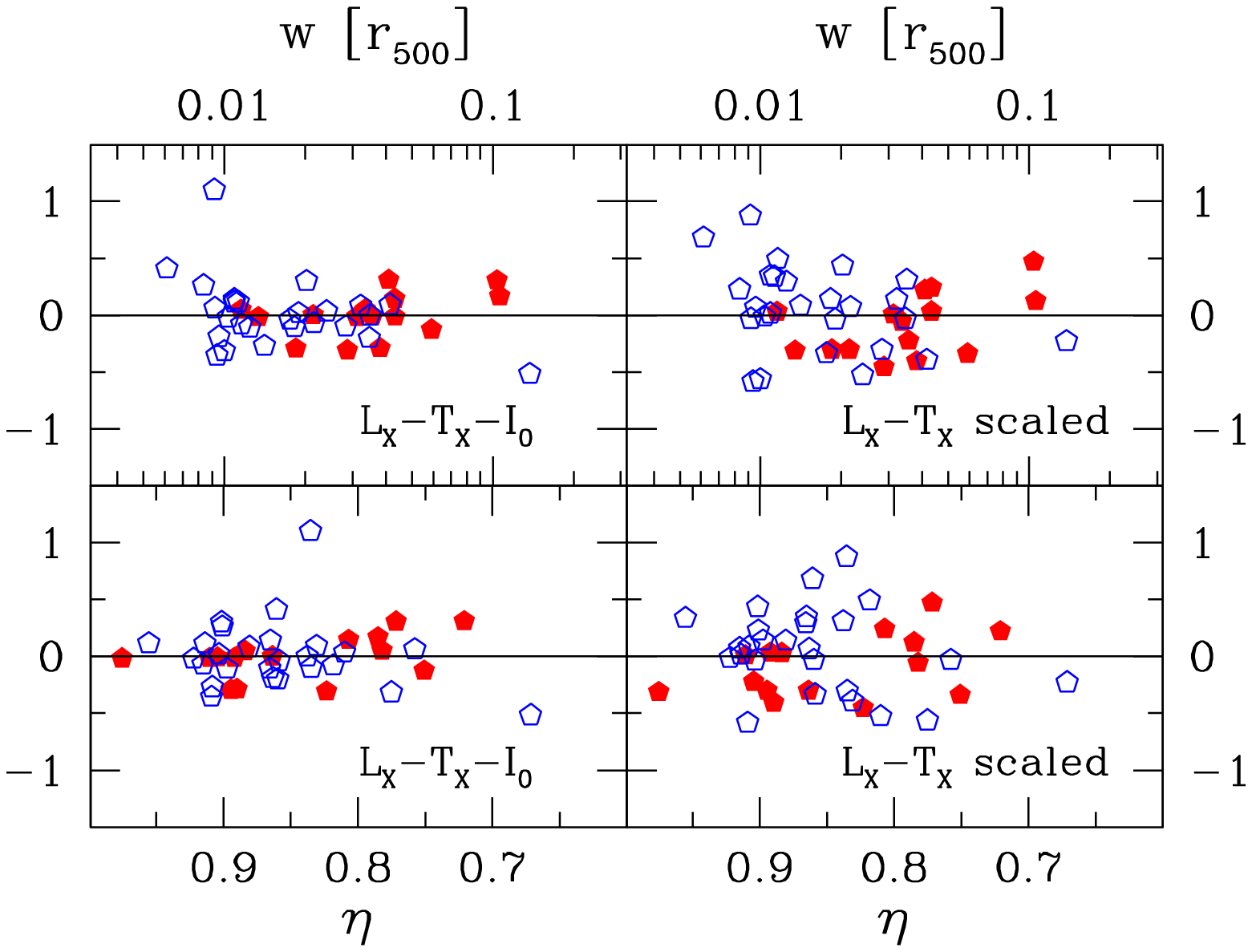}\end{minipage}
   \end{center}}
   \figcaption{\label{fig:diffLx} 
Logarithmic (base-$e$) deviations of clusters from best-fit \LXfive--\Tx--\Io\  scaling relation ({\it left})  and best-fit \LXfive--\Tx\ relation ({\it right}) versus centroid variation $w$ ({\it top}) and axial ratio $\eta$ ({\it bottom}). Open and filled markers are CC and NCC clusters, respectively. Centroid variations are given in units of \rfive.  
  }
\end{inlinefigure}

\begin{inlinefigure}
   {
   \begin{center}
   \epsfxsize=9.cm
   \begin{minipage}{\epsfxsize}\epsffile{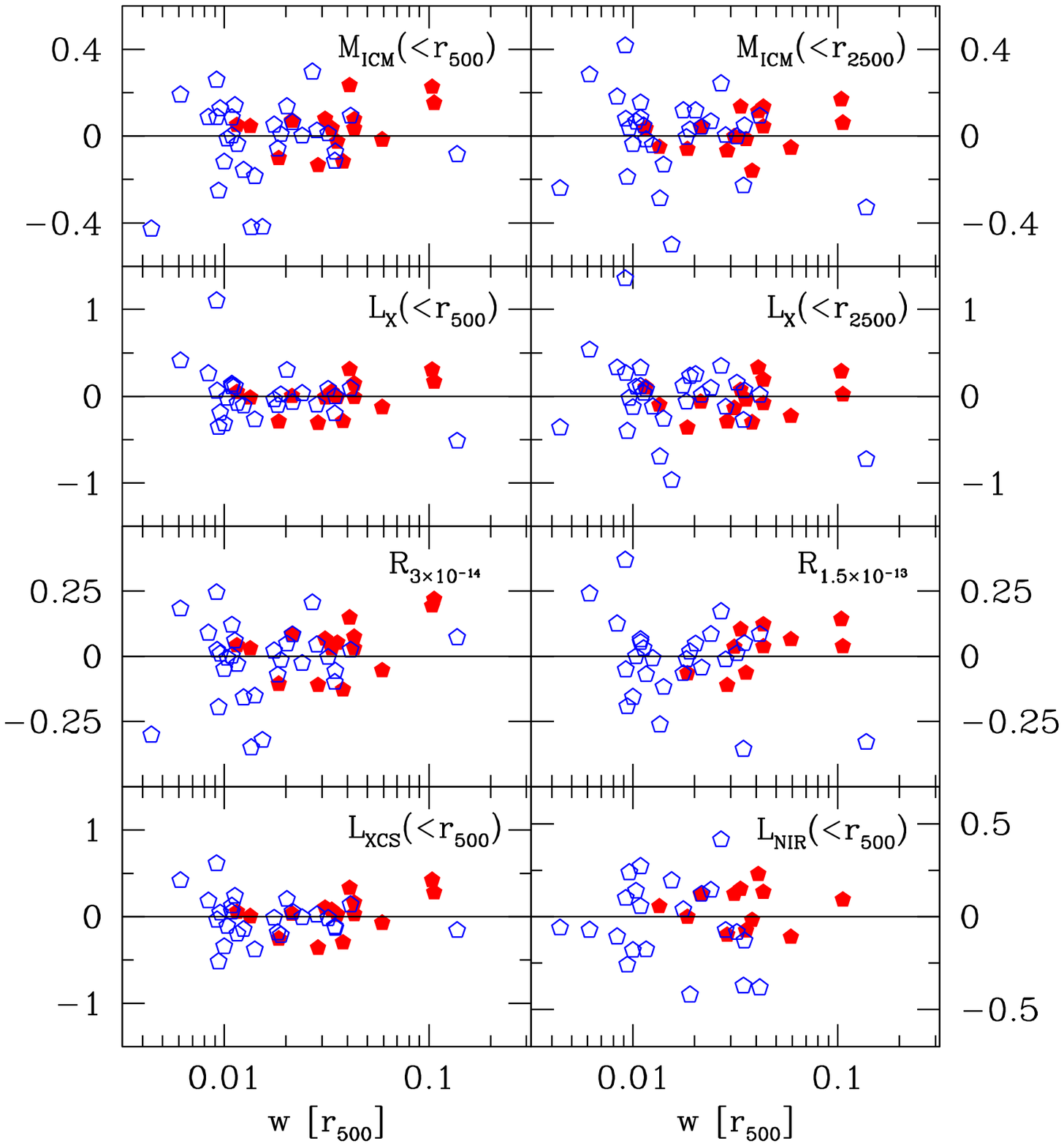}\end{minipage}
   \end{center}}
   \figcaption{\label{fig:diffobs} 
Logarithmic (base-$e$) deviations of clusters from best fit scaling relations versus centroid variation $w$ for each of the seven X-ray observable--temperature--brightness scaling relations and the NIR luminosity--temperature--brightness relation. Open and filled markers are CC and NCC clusters, respectively. Centroid variations are given in units of \rfive.   
  }
\end{inlinefigure}

\subsection{Substructure and Scaling Relations: Individual Cluster Relations}
\label{sec:subsscale}

We now look for merger-related structural variations in all scaling relations.  Figure~\ref{fig:diffobs} shows the natural logarithmic deviation of each data point from the best fit scaling relation for our sample plotted versus centroid variation, and Figure~\ref{fig:diffobs_eta} shows the deviation versus axial ratio. The most obvious feature of these data is the semi-separation of CC and NCC clusters by substructure indicator; i.e., the CC clusters tend to have smaller centroid variations than the NCC clusters. This relationship between cool core status and axial ratio is not as striking, but is still present.

\begin{inlinefigure}
   {
   \begin{center}
   \epsfxsize=9.cm
   \begin{minipage}{\epsfxsize}\epsffile{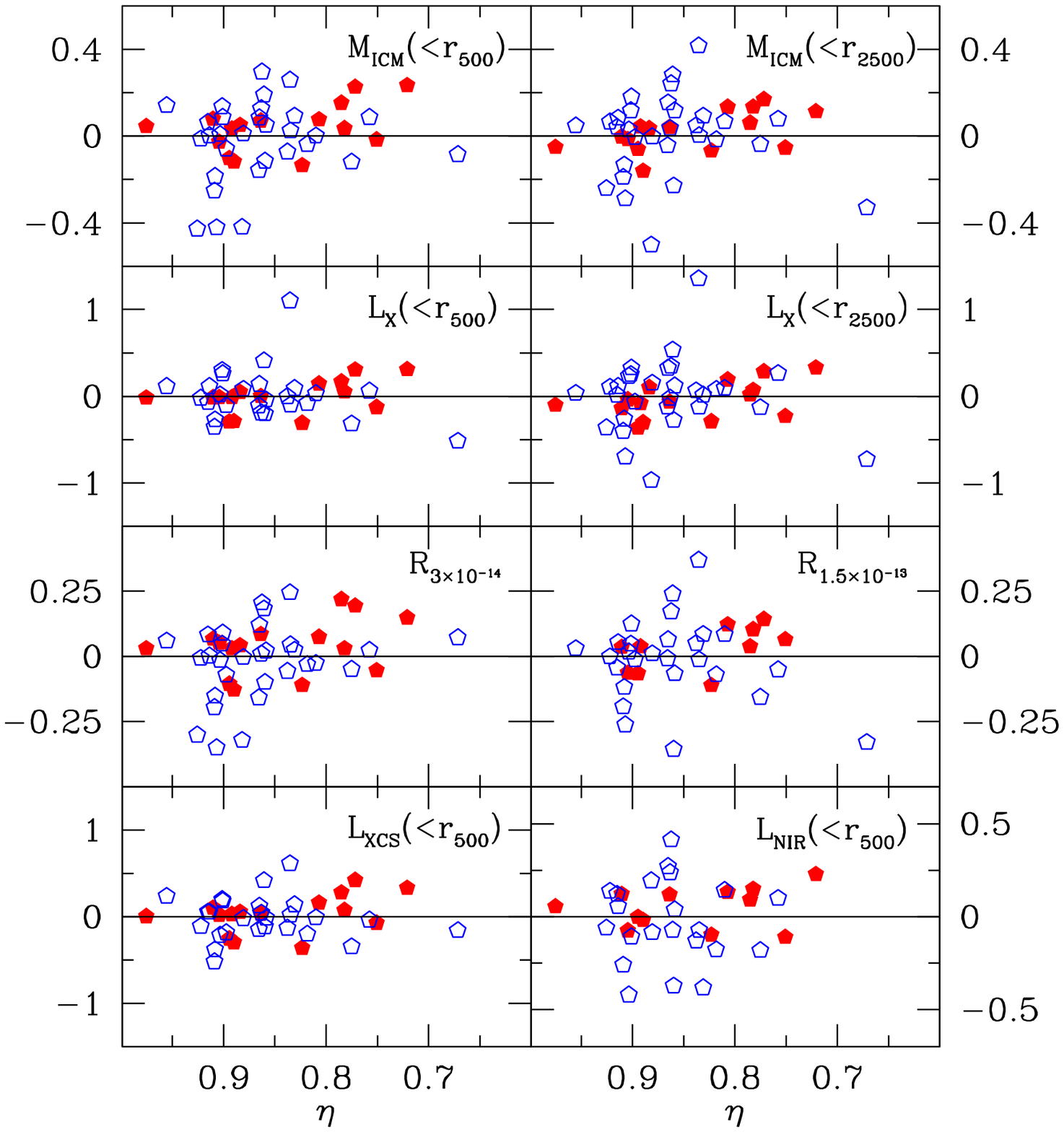}\end{minipage}
   \end{center}}
   \figcaption{\label{fig:diffobs_eta} 
Logarithmic (base-$e$) deviations of clusters from best fit scaling relations versus axial ratio $\eta$ for each of the seven X-ray observable--temperature--brightness scaling relations and the NIR luminosity--temperature--brightness relation. Open and filled markers are CC and NCC clusters, respectively. Axial ratio increases to the left, i.e., ellipticity increases to the right.  
  }
\end{inlinefigure}

Figs.~\ref{fig:diffobs} and \ref{fig:diffobs_eta} show no  qualitative evidence for an increase in scatter in clusters with higher substructure, and indeed suggest greater scatter in clusters with less substructure. There may be a trend for the most irregular clusters to lie above the scaling relation, but the small number of clusters involved (2--3) makes this very uncertain. Cluster deviations from scaling relations versus \Pone and \Ptwo\ (not plotted here) likewise show no suggestion of higher scatter in clusters with greater substructure.

To quantitatively address the issue, we measure the intrinsic scatter in the scaling relations for subsamples grouped by  centroid variation, axial ratio, \Pone, and \Ptwo. As before, we add a value for the intrinsic scatter \sint\ in quadrature to the uncertainty in the observable due to measurement and temperature uncertainty, and find the value of \sint\ that results in a reduced $\chi^2$ value of unity for each scaling relation. We express \sint\ in units of natural logarithm of the observable, i.e., the units of the vertical axes in Figs.~\ref{fig:diffobs} and \ref{fig:diffobs_eta}.

We split the sample into two subsamples for each substructure measure; the split point for each is chosen to include roughly half the clusters in the sample.   Specifically, we split the sample at $w=0.02$, $\eta=0.875$, \Pone\ $=300$, and \Ptwo\ $=20$. Note that for the axial ratio $\eta$, a higher value corresponds to a more regular (spherical) cluster, whereas for the other substructure measures a higher value is, roughly speaking, a messier cluster. 
Table~\ref{tab:subsplit} contains the results; for ease of interpretation, a graphical representation of the same data is shown in Fig.~\ref{fig:scatterbar}. As in Table~\ref{tab:ccsplit}, we give the percent significance level at which equality of subsample variances is rejected.

Broadly speaking, we find greater scatter in clusters with less substructure; with 11 relations and four substructure measures, we find significantly (i.e., same scatter rejected at $>68\%$ level) greater scatter in low substructure clusters in 13 cases, and in high substructure clusters in only four. This is remarkable, although it is not surprising in light of our earlier result (\S~\ref{sec:cores}) that there is greater scaling relation scatter in the CC population than in the NCC population.  One might suspect that a few outliers or a poor choice of splitting values could cause an apparent increase in scatter in lower substructure clusters, but an examination of Figure~\ref{fig:diffobs} does not support this. 
Indeed, while the scatter measurements suggest that clusters with higher ellipticity may have greater scatter, Figure~\ref{fig:diffobs_eta} suggests that this is a result of a few outliers, and does not constitute a general trend to higher scatter in more elliptical clusters.

\begin{inlinefigure}
   {
   \begin{center}
   \epsfxsize=8.cm
   \begin{minipage}{\epsfxsize}\epsffile{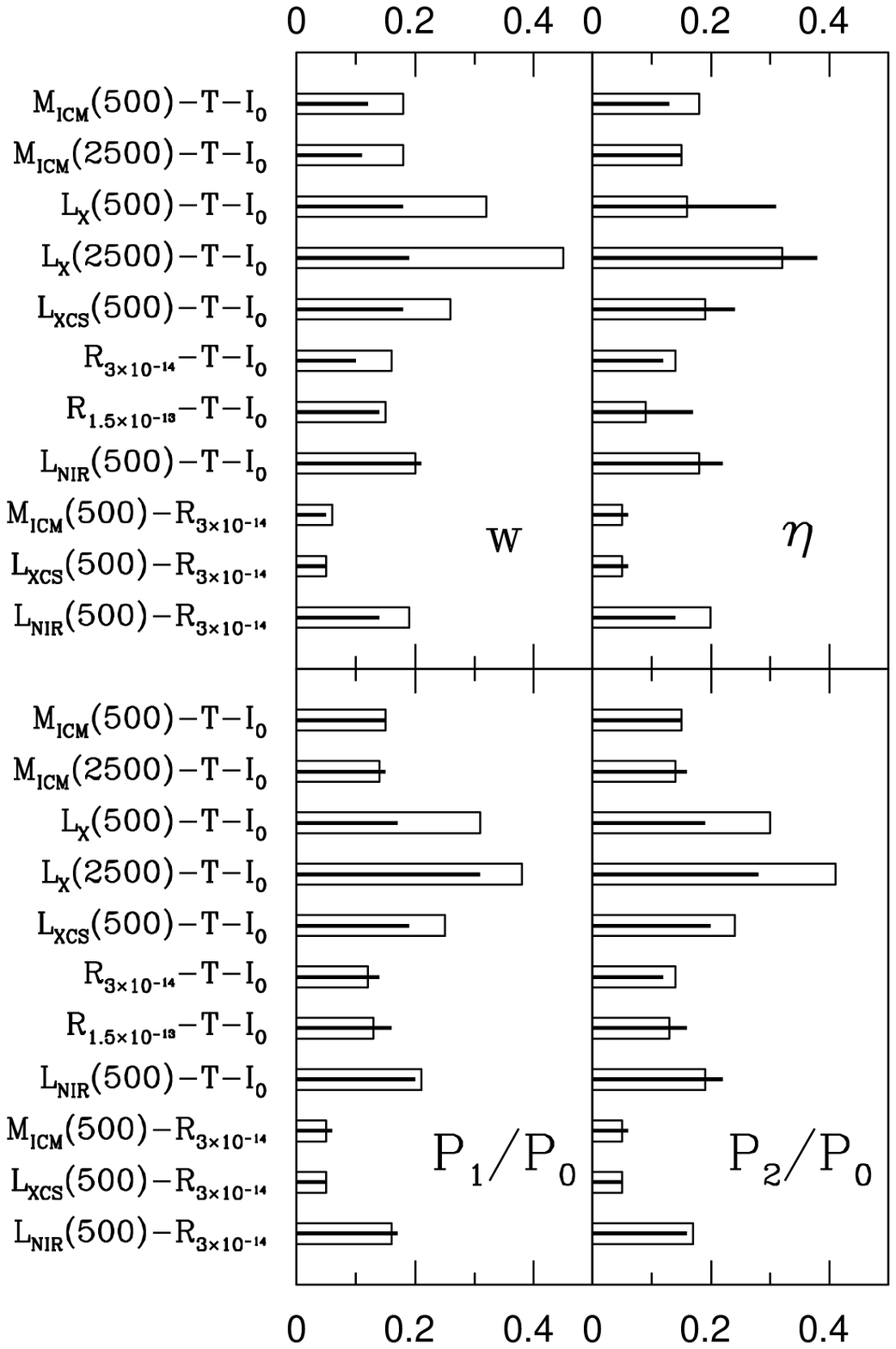}\end{minipage}
   \end{center}}
   \figcaption{\label{fig:scatterbar} 
Graphical representation of the observational sample data in Table~\ref{tab:subsplit}. Open and closed bars are for low-substructure and high-substructure populations, respectively.
  }
\end{inlinefigure}

One way to minimize the effects of the cool cores is to examine observables that are least affected by the cool cores.  We examine deviation in \RI\ from the \Rthree--\Micmfive\ scaling relation that was plotted in Figure~\ref{fig:RM500}.  The emission-weighted mean temperature does not appear in this analysis, and these two observables are very insensitive to the core structure of the ICM.  Deviations  from this scaling relation are plotted versus centroid variation in Figure~\ref{fig:diff_RM500}. Total scatter in this relation is smaller than for any of the observable--temperature relations, providing another indication that is it the core structure of clusters which varies most significantly within the population.  The cluster deviation versus $w$ provides some indication of higher scatter at lower substructure. We do see the same suggestion of a preferential boosting above the scaling relation for a few clusters at high $w$ as for the observable--temperature relations discussed above.  We also examine scaling relations constructed from \Rthree-\LXCSfive\ and \Rthree-\LNIRfive; these, too, show little if any separation between CC and NCC populations, and exhibit little evidence for different scatter for populations with different levels of substructure.

In summary, the evidence clearly does not support our naive expectation that clusters with more substructure should exhibit higher scatter than their more relaxed counterparts.

\begin{inlinefigure}
   {
   \begin{center}
   \epsfxsize=8.cm
   \begin{minipage}{\epsfxsize}\epsffile{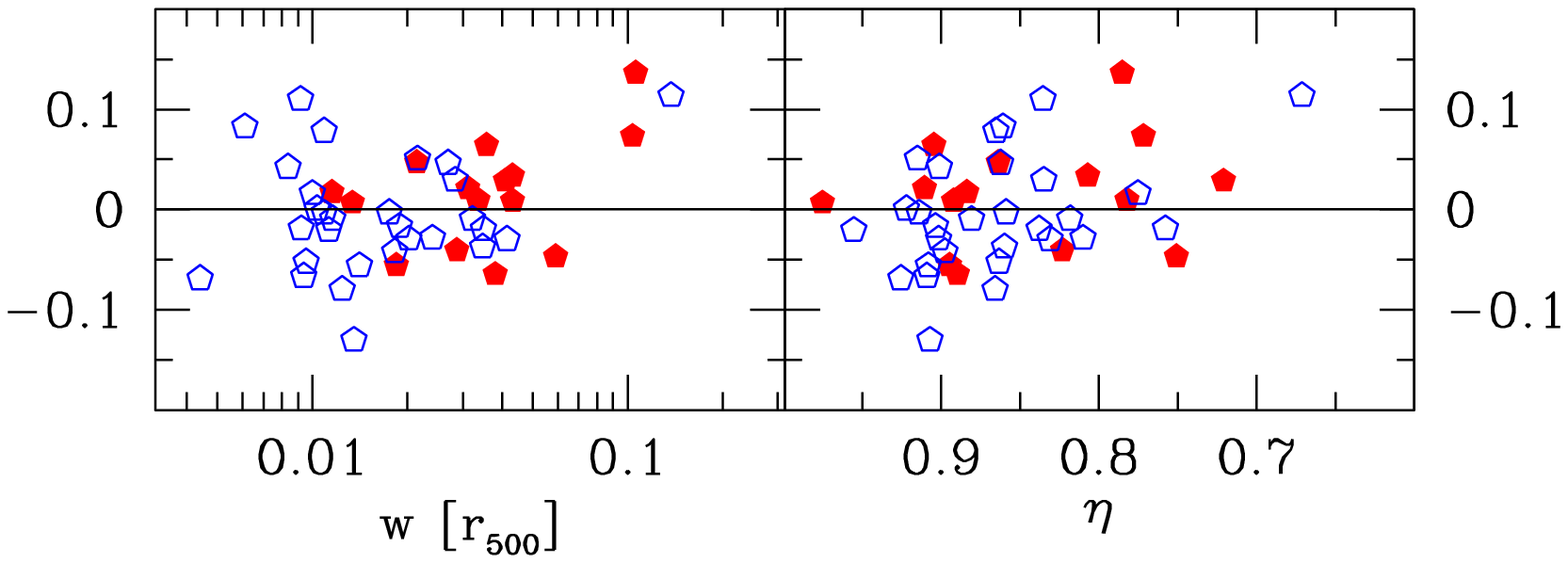}\end{minipage}
   \end{center}}
   \figcaption{\label{fig:diff_RM500} 
   As in Fig.~\ref{fig:diffobs}, but for deviation in \RI\ from the \Rthree--\Micmfive\ scaling relation.
     }
\end{inlinefigure}

\subsection{Substructure and Multiple Scaling Relations}

Although we have shown that the clusters with the most substructure do not preferentially deviate from individual scaling relations, it is possible that within the hyperspace defined by our broad range of observables these merging systems may have a tendency to lie somewhat further from the general population.  We examine this possibility by combining cluster deviations from all the scaling relations and probing for greater combined deviations in systems with the most substructure.  To do this we assume that the cluster behavior about  a scaling relation is a probabilistic indication of the consistency of that cluster with the typical structure of the population.  Specifically, we assume that the probability of finding a cluster at a given deviation is described by a Gaussian in log space centered on the relation with a full-width at half-maximum equal to $2.354\sigma_{\log {\mathcal{O}}}$, where $\sigma_{\log {\mathcal{O}}}$ is the intrinsic scatter of the population about the scaling relation.   That is, we calculate 
\begin{equation}
\label{eq:Gauss}
P_{\mathcal{O}}  = \frac{1}{\sqrt{2\pi}\sigma_{\log {\mathcal{O}}}} \exp \left[ -\frac{1}{2} \left( \frac{\log {\mathcal{O}} - \log {\mathcal{O}}_{\rm fit}(T_{\rm X}) }{\sigma_{\log {\mathcal{O}}}} \right)^2 \right]
\end{equation}
for an observable $\mathcal{O}$. We also calculate $P_{\rm total}$, a measure of the significance of deviation from multiple scaling relations together, by multiplying the individual probability amplitudes. If a cluster deviates slightly from three individual relations, for example, this should be reflected in $P_{\rm total}$.  This approach assumes that the cluster behavior about each scaling relation is an independent indicator of its deviation from the whole population.

We find no correlation between deviation from scaling relations and substructure for any individual \OT\ scaling relation, or for all observables measured together. The three highest substructure clusters noted in the previous discussion do indeed have low total probability densities, as expected, but this is also true of several other clusters over the entire range of substructure in our cluster sample. We conclude that there is no readily discernible relationship between substructure and deviation from scaling relations by individual clusters.


\subsection{Hydrodynamical Cluster Simulations}
\label{sec:sims}

Although simulated clusters do not exhibit the full complexity of real clusters, carrying out our analysis on a controlled sample of well-understood systems is an important component of our work.
We use a simulated cluster ensemble consisting of 45 smoothed particle hydrodynamics simulations evolved in a $\Lambda$CDM cosmology; details of the techniques used and of this particular sample can be found in \citet{bialek01,bialek02}. These simulations have resolution sufficient to exhibit cluster merger features such as cold fronts \citep{bialek02}. They do not include any ICM cooling mechanism, and hence cool cores will be absent; this provides a good opportunity to examine the results that would be expected in our observational sample if we were able to completely remove the cool core effects.  Lack of resolution and incomplete physical modeling make the central, core regions of simulated clusters unreliable, and so we choose to examine scaling relations that are less sensitive to these core regions.

\begin{inlinefigure}
   {
   \begin{center}
   \epsfxsize=8.cm
   \begin{minipage}{\epsfxsize}\epsffile{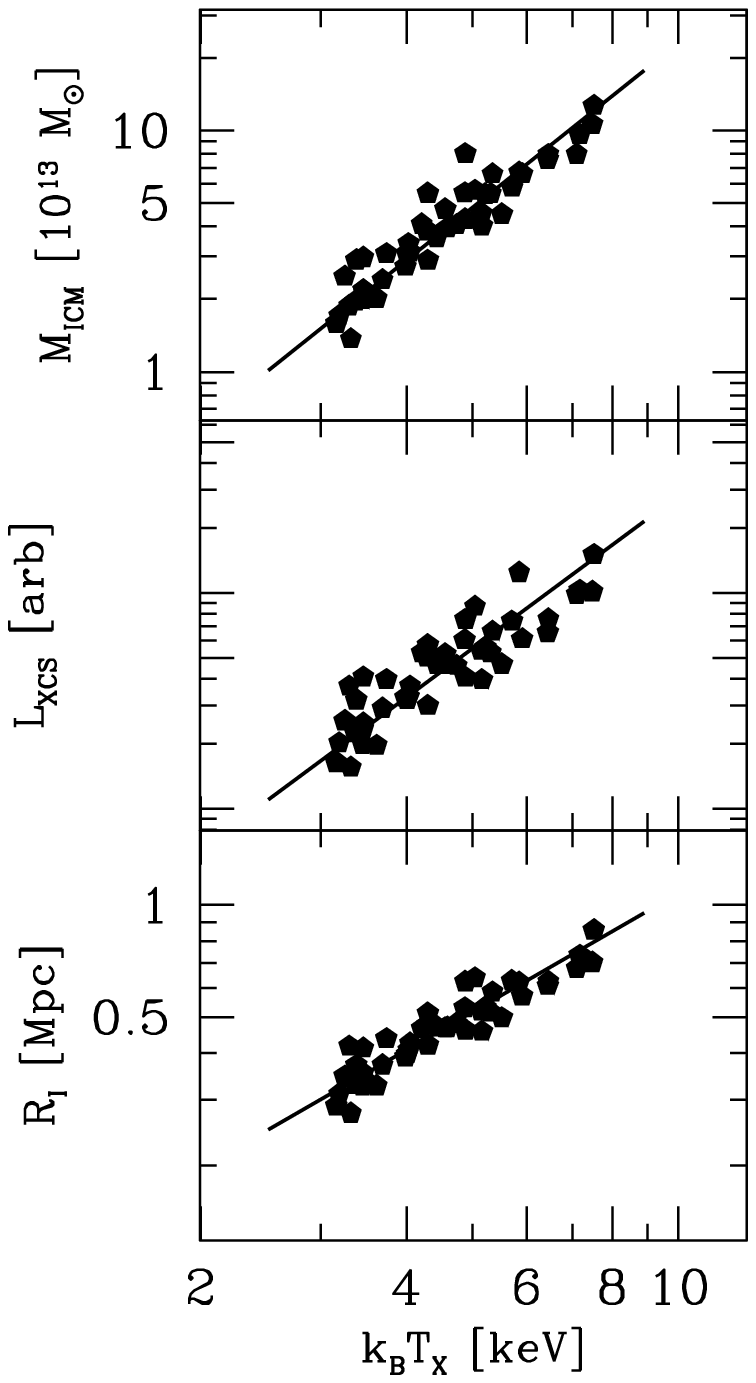}\end{minipage}
   \end{center}}
   \figcaption{\label{fig:scalingsim} 
   Simulated scaling relations for \Micm\ within \rtwo, core-subtracted X-ray luminosity projected within \rfive, and isophotal size for an isophote of  $1\times10^{-3}$ counts \ps\  arcmin$^{-2}$, versus \Tx.  
     }
\end{inlinefigure}

We examine the projected \ROSAT-band (0.5--2.0 keV) core-subtracted luminosity within \rfive; that is, without the luminosity projected within $0.20r_{500}$, as in our observed cluster sample. 
We derive the scaling relation for \Micmfive\ using the actual simulation data (i.e., not calculating \Micm\ from mock observations). We measure the isophotal size in the \ROSAT\ band corresponding to an isophote of $1\times10^{-3}$ counts \ps\ arcmin$^{-2}$; this instrumental isophote leads to isophotal sizes that approximately match the normalization of the observed \Rthree--\Tx\ relation.
 This isophote generally lies well outside the core of the simulated clusters.
 
 Scaling relations for these three observables are shown in Figure~\ref{fig:scalingsim}.
As we did for the observed scaling relations, in Figure~\ref{fig:diffsim} we plot the difference between each cluster and the best-fit relations versus the $w$ and $\eta$ substructure indicators. There is some suggestion here of a trend toward greater scatter at higher substructure. To quantify this, we calculate the intrinsic scatter as was done for the observations, both for the entire sample and for subsamples of roughly equal size; values of \sint\ are shown in Table~\ref{tab:subsplit}. 

\begin{inlinefigure}
   {
   \begin{center}
   \epsfxsize=8.cm
   \begin{minipage}{\epsfxsize}\epsffile{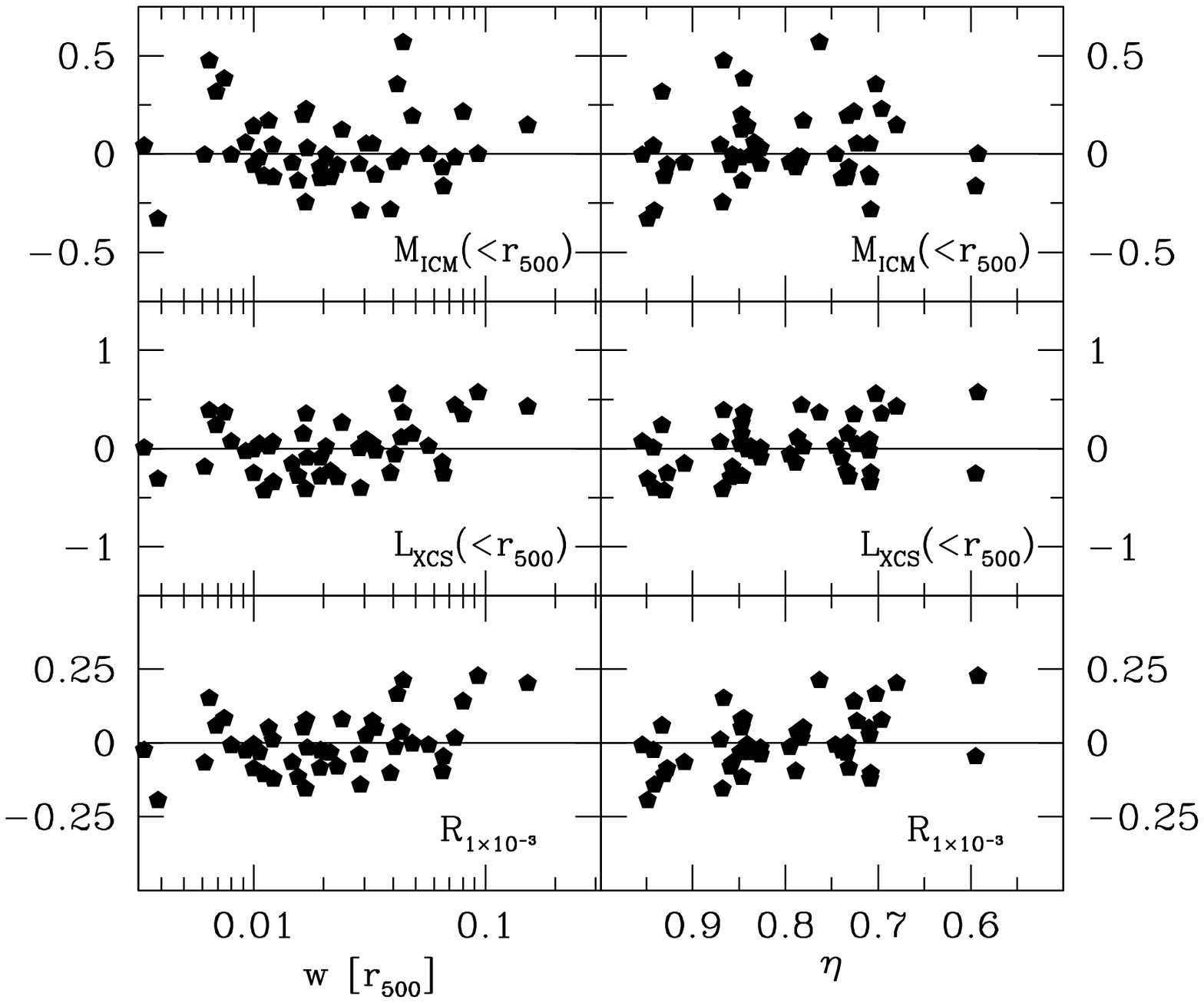}\end{minipage}
   \end{center}}
   \figcaption{\label{fig:diffsim} 
   Deviation in base-$e$ logarithm of simulated clusters and from best-fit relations plotted versus centroid variation $w$ and axial ratio $\eta$ for each of the three simulated X-ray scaling relations. Centroid variation is given in units of \rfive.
     }
\end{inlinefigure}

Looking at all four substructure measures, there is a tendency toward lower scatter in clusters with less substructure. However, in general the quantitative differences are rather small; significantly ($>68\%$ significance) different subsample scatter is found in only two relations, and greater scatter in low substructure clusters in only one.  We also calculate  the probability amplitude for individual clusters by combining information from all scaling relations, as we did for the observations. This approach provides no suggestion of greater or smaller deviation for clusters with less substructure.  Thus, while the simulations do not show evidence for greater scatter in  clusters with more substructure, they also do not show the tendency toward higher scatter in less messy clusters that we see in the observed cluster sample.

The scatter we measure in the simulated sample can be compared directly to the intrinsic scatter in the observed samples.  Because of the lack of radiative cooling in the simulations, it is likely best to compare the simulation scatter to the cool core ``corrected" observations (i.e., the scatter in the \OTI\ relations).  Agreement in the scatter of simulated and observed scaling relations would serve as one more indication \citep[along with the slope and normalizations;][]{bialek01} that the simulations are an accurate representation of real clusters.  Interestingly, the scatter is greater for the simulations in mass (0.20:0.14) and luminosity (0.27:0.21), and smaller in isophotal size (0.10:0.13).  The comparison is not entirely appropriate in the case of the ICM mass, because in the simulations we use the actual three dimensional ICM mass measured within \rfive, and for the observations we calculate this through a deprojection.  The bottom line is that although these simulations with early and uniform preheating do a good job of reproducing the observed slopes and amplitudes of scaling relations, there is still clearly work to be done to match the scatter in observed scaling relations.  The range of missing physics in the simulations includes radiative cooling, conduction, magnetic fields, feedback from AGN, galaxy formation, etc., and all of these must be considered together with the effects of finite spatial resolution.  It is quite interesting to consider that some ingredients currently missing from simulations could actually reduce the scaling relation scatter---that is, reduce the cluster to cluster structural variations at a given mass.

\subsection{Summary of Substructure Results}
\label{sec:subsum}

We have shown that clusters with greater substructure do not preferentially lie significantly farther from scaling relations than clusters with less substructure. In fact, in observed  clusters, there is a tendency toward greater scatter about scaling relations in apparently more relaxed clusters. These findings contradict the naive expectation that cluster structure is greatly disturbed by merger events and so should increase scaling relation scatter. 

Comparison of the observational and simulated cluster samples suggests that the greater scatter in more apparently relaxed clusters must be the result of processes not present in the simulations.
Cool core-related phenomena are clearly the prime candidate for the higher scatter seen in observed clusters with less substructure, especially as our results show unquestionably higher scatter in CC clusters when compared to NCC clusters. AGN activity may also contribute, as AGNs occur frequently in the cluster population and can produce radio cavities with associated energies of at least $\sim 10^{60}$ erg \citep{birzan04}. 

However, even the simulations show only weak evidence of higher scatter in clusters with more substructure. We must conclude that either there are mechanisms which introduce a range of structural variations into apparently relaxed clusters, or that mergers simply do not perturb cluster structure to the extent expected and suggested by simulations of isolated clusters \citep{ricker01,randall02}.

It may be suggested that the use of archival temperatures from different sources may introduce scatter or otherwise hide merger or cooling effects. Redoing the \LXfive\ portion of our analysis with 43 of 45 temperatures taken from a single published source \citep{dwhite00} demonstrates that this is not the case; while quantitative scatter measurements can be sensitive to outliers, we find no evidence that the qualitative trends we report are affected by moderate shifts in cluster temperatures.

\section{Conclusions}
\label{sec:concl}

We examine the relationship between substructure and cool cores in galaxy clusters and the scatter about X-ray and near-IR scaling relations using emission-weighted, non-spatially resolved, non-cooling corrected temperatures.  We separate clusters into CC and NCC subsamples according to their central cooling times, and we quantify substructure using the centroid variation, axial ratio, and power ratios \Pone\ and \Ptwo.  With these tools and a wide range of crude cluster observables and uncertainties, we examine a large number of galaxy cluster scaling relations in an approximately X--ray flux limited sample of 45 clusters.   Our primary findings are:

\begin{enumerate}

\item There is an offset between CC and NCC populations on all observable--temperature scaling relations that we examine.  This separation is partly due to an emission weighted mean temperature bias at around the 7\% level.  The offset must also be due to differences in the core structure of the two subsamples, because those scaling relations that are most sensitive to the cluster core show the largest offsets.   This offset is not driven by recent mergers, because scaling relations involving observables that are sensitive primarily to the outer structure of clusters show larger scatter in the CC population.

\item We show that the central X--ray surface brightness can be used to characterize the ``strength" of cool cores, and that introducing it as a third parameter in observable--temperature scaling relations  greatly reduces the scatter about those relations. Thus, the central surface brightness provides a tool for studying the evolution of cluster scaling relations in a manner less sensitive to any change in the fraction of cool core clusters with redshift.  In addition, the small scaling relation scatter when using the surface brightness means that crude cluster observables like the X--ray luminosity and temperature can provide more accurate virial mass estimates than are obtained without the third parameter.

\item Parameterized in terms of central surface brightness \Io, we find that the emission weighted mean temperature bias correction factor is proportional to $I_0^{(0.03 \pm 0.01)}$. Given the factor of $\sim$500 variation in \Io\  within our sample, this implies a maximal differential correction across our sample of $\sim20$\%.  Using the median \Io\ for the CC and NCC populations, we calculate an average CC temperature bias factor of $1.07 \pm 0.02$ for this sample.

\item We find that although CC clusters tend to exhibit less morphological substructure, 
they exhibit at least as much scatter as (and often more than) the NCC clusters.  Thus, structural variations among CC clusters are at least as large as the structural variations among NCC clusters.  This result has important implications for analyses which rely on the presence of a cool core to indicate that a cluster is relaxed. 

\item Clusters with greater morphological substructure do not exhibit more scatter about scaling relations than clusters with lower substructure. In fact, we observe  a trend toward higher scatter in clusters with less substructure. This may be partially due to the structural variations within cool cores, which are typically found in clusters that exhibit less substructure; however, even after using central surface brightness to reduce the systematic cool core effects, we find that the clusters with less substructure exhibit as much or more scatter as those with more substructure. 

\item  The differences between low and high morphological substructure clusters are modest in our study of hydrodynamical cluster simulations without cooling;  there is only a weak indication that those clusters with higher substructure exhibit higher scatter.  

\item As in the purely X--ray scaling relations, there is no relationship between deviation from the \LNIR--\Tx\ relation and cluster substructure.  However, in the NIR relation there is a negative temperature scale factor required to align the CC and NCC populations, and a lack of any dependence on \Io\ in a constructed \LNIR--\Tx--\Io\ relation.   Because we know that emission-weighted mean temperatures are biased by the cool core gas in CC clusters, this suggests a difference between galaxy populations in CC and NCC clusters. A detailed study of differences in NIR properties of the galaxy population in CC and NCC clusters would clearly be interesting.

\end{enumerate}

Together, these results from studies of real and simulated clusters indicate that cool core related phenomena (such as radiative cooling, AGN activity, and entropy injection at an earlier epoch), and not cluster merging, are the primary sources of scatter in scaling relations.   Perhaps it should come as no surprise that X--ray observables that arise from emission, which is sensitive to the square of ICM density, are most perturbed by the detailed structure of the cluster core.  However, the lack of a strong substructure related enhancement of scatter in scaling relations without sensitivity to the cluster core and in hydrodynamical simulations of clusters, is surprising.  It suggests that perhaps all clusters retain departures from equilibrium at a significant enough level that even recent mergers do not perturb their structure enough to make them appear unusual.  A young population with relaxation timescales that are comparable to the time since the last major merger would presumably exhibit this kind of behavior.

Reconciling these observations and our conclusions with high resolution hydrodynamical mergers of idealized clusters \citep[e.g.,][]{ricker01} requires a rarity of such large-scale mergers or perhaps additional physics within the ICM that suppresses the boosts.  Note that correlated excursions in luminosity and temperature during a merger will not suffice as an explanation, because we have shown using 8 different observable--temperature scaling relations that there is no strong relationship between substructure and scaling relation scatter.  We have demonstrated here that lower resolution hydrodynamical simulations simply do not produce large deviations from the general cluster population even when there is evidence for significant substructure, either in individual clusters or in the high-substructure cluster population as a whole.  Whatever the explanation, it is clear from reasonably large samples of real clusters that there are no outliers on the scale of those predicted by the high resolution, idealized cluster merger simulations.

Our result may be consistent with the scenario where NCC clusters evolve to become CC clusters in the absence of major mergers \citep[e.g.,][]{ota05}.  However, the larger scatter we observe for CC clusters in all scaling relations raises important questions.  In particular, even after using the central surface brightness to correct for CC effects, we still observe higher scatter about the CC relations (see Table~\ref{tab:ccsplit}).  Moreover, we see the larger scatter in CC clusters even in scaling relations that involve observables that are not sensitive to the core structure (i.e., faint isophotal size, ICM mass within \rfive, and NIR light).  Thus, if CC clusters evolve from NCC clusters because of an absence of mergers, then the observations require some other source of scatter or variation in cluster structure to be present throughout the cluster virial region. 
 
Alternatively, cool cores may arise through a scenario that is driven by something other than the recent merger history of the cluster.   \citet{mccarthy04} suggest that variations in entropy injection into the intracluster medium could determine whether or not a cool core forms.  This varied entropy injection would also contribute to structural variations or scatter in scaling relations.  Because our cluster sample indicates that it is the CC clusters which exhibit the highest scatter around scaling relations (even those relations with little core sensitivity), it seems likely that this or some similar, non-merger driven scenario is responsible for the presence or absence of cool cores in clusters. Within this scenario the tendency for cool core clusters to exhibit less morphological substructure would be primarly due to the effects of the often dominant X-ray bright core, which would tend to bias axial ratios high and centroid variations low.  That is, in cool core clusters a morphological substructure indicator is in large part reflecting the characteristics of the bright, symmetric core.

It will be quite interesting to return to this question of the dual nature of galaxy clusters---youthful as indicated by the high frequency of morphological substructure, yet strikingly regular as indicated by scaling relations---with new tools and larger samples extending over a wider range of redshift.  Of particular interest will be the additional leverage afforded by the new generation of high signal to noise Sunyaev-Zel'dovich effect observations, which should be dramatically less core-sensitive than X--ray observations.  With the tens of thousands of clusters expected in dedicated surveys, it should be possible to quantify with high significance any subtle, merger related trends that may be present.

\acknowledgments

This work was supported in part by the NASA Long Term Space Astrophysics award NAG 5-11415. This publication makes use of data products from the Two Micron All Sky Survey, which is a joint project of the University of Massachusetts and the IPAC/Caltech, funded by NASA and the NSF.

\bibliographystyle{apj}
\bibliography{cosmology}

\end{document}